\documentclass[10pt,fleqn]{article}

\usepackage{color}

\newcommand{\bed}{\[}
\newcommand{\eed}{\]}
\newcommand{\beq}{\begin{equation}}
\newcommand{\eeq}{\end{equation}}
\newcommand{\beqa}{\begin{eqnarray}}
\newcommand{\eeqa}{\end{eqnarray}}
\newcommand{\ket} [1] {\vert #1 \rangle}
\newcommand{\bra} [1] {\langle #1 \vert}

\newcommand{\mean}[1]{\langle #1 \rangle}
\newcommand{\gras}[1]{\bold{#1}}

\newcommand{\tr}{\mathop{\mathrm{tr}}}

\newcommand{\be}{\begin{eqnarray}}
\newcommand{\ee}{\end{eqnarray}}
\newcommand{\bea}{\begin{eqnarray}}
\newcommand{\eea}{\end{eqnarray}}
\newcommand{\bma}{\begin{subequations}}
\newcommand{\ema}{\end{subequations}}

\renewcommand{\>}{\rangle}

\newtheorem{lemma}{Lemma}

\usepackage{graphicx}
\usepackage[mathscr]{eucal}
\usepackage{amsmath}
\usepackage{amssymb}
\usepackage{epstopdf}
\usepackage{epsfig}
\usepackage{wrapfig}
\def\one{\ensuremath{\hbox{$\mathrm I$\kern-.6em$\mathrm 1$}}}
\def\tr{ \mbox{tr}}

\begin{document}

\title{Thermal States of Anyonic Systems}

\author{
S. Iblisdir$^{1}$, D. P\'erez-Garc\'ia$^{2}$, M. Aguado$^{3}$, J. Pachos$^{4}$\\
\normalsize \it
$^1$Dpt. Estructura i Constituents de la Materia, \\
\normalsize \it
Universitat Barcelona, 08028 Barcelona, Spain\\
\normalsize \it
$^2$Dpt. An\'alisis Matem\'atico, \\
\normalsize \it
Universitad Complutense de Madrid, 28040 Madrid, Spain\\
\normalsize \it
$^3$ Max Planck Institut f\"ur Quantenoptik, \\
\normalsize \it
Garching D-85748, Germany\\
\normalsize \it
$^4$ School of Physics and Astronomy, \\
\normalsize \it
University of Leeds, Leeds LS2 9JT, United Kingdom
}

\date{}

\maketitle

\begin{abstract}
%

A study of the thermal properties of two-dimensional topological lattice models  is presented. This work is relevant to assess the usefulness of these systems as a quantum memory. For our purposes, we use the topological mutual information $I_{\mathrm{topo}}$ as a ``topological order parameter''. For Abelian models, we show how $I_{\mathrm{topo}}$ depends on the thermal topological charge probability distribution. More generally, we present a conjecture that $I_{\mathrm{topo}}$ can (asymptotically) be written as a Kullback-Leitner distance between this probability distribution and that induced by the quantum dimensions of the model at hand. We also explain why  $I_{\mathrm{topo}}$ is more suitable for our purposes than the more familiar entanglement entropy $S_{\mathrm{topo}}$. A scaling law, encoding the interplay of volume and temperature effects, as well as different limit procedures, are derived in detail. A non-Abelian model is next analysed and similar results are found. Finally, we also consider, in the case of a one-plaquette toric code, an environment model giving rise to a simulation of thermal effects in time. 
\end{abstract}

\section{Introduction}\label{sec:intro}

Quantum mechanics has significantly shaped our current understanding of
condensed matter systems. It has proven to be deeply insightful when aiming
to explain, say, transport of charged (quasi-) particles in semiconductors,
magnetism in metallic alloys or cohesive properties of solids
\cite{ashcroft}. But when it comes to macroscopic systems that exhibit a
highly non-classical behavior, many fundamental issues are still poorly
understood. An example of such systems is a fractional quantum Hall effect
sample (FQHE), i.e. a specific  two-dimensional electron gas subject to a
strong perpendicular magnetic field. There the transverse conductivity
appears in plateaux at fractional values of the filling factor, much in
contrast to what one would expect classically~\cite{FQHE}. Besides their fundamental interest, 
the exotic phases exhibited by these systems are important in that they may lead to new 
technological applications. Indeed, they could allow for intrinsically fault-tolerant quantum computation~\cite{Kit97,tqc:review}. These phases are not separated by a symmetry whose absence or presence can be detected by a~\emph{local} order parameter. Rather, it has been realized that they are associated with topological order.

In the effort towards understanding this notion, an important
contribution has been the research towards lattice spin models. Certain such
systems are exactly solvable~\cite{Kit97,Levin-Wen05,Douc03,bombin}, in the
sense that their low energy sectors can be analytically determined. Although
the corresponding interactions are local and frustration-free, they possess
the essential properties of topologically ordered systems: (i) when the
system is defined on a surface with non-trivial topology, the ground state
has a non-local degeneracy that cannot be detected locally, (ii) excitations
have exotic statistics. These features turn these models into an interesting
alternative to FQHE systems, when it comes to study topological order.
Initially, topological spin lattice systems were proposed as a reliable
quantum memory~\cite{Kit97,Den01,Wan02} and  some non-Abelian versions were
shown to allow for universal fault-tolerant quantum
computation~\cite{mochon}. Quantum information would be encoded in non-local
degrees of freedom of such systems and would therefore be immune to local
perturbations. Fault-tolerant quantum computation could be performed by
creating excitations (initialization), braiding them (unitary evolution),
and fusing them back together (read-out) \cite{tqc:review,brennen}.

In this work, which is partly an extension of \cite{Iblisdir}, we focus on an entropic order parameter, 
the topological mutual information, $I_{\textrm{topo}}$, and use it to study how topological
matter behaves in the presence of temperature. For concreteness, we restrict
our analysis to two paradigmatic models: the toric code and the $D(S_3)$
``superconductor'' \cite{Kit97,Pres:chap9}.  We will see that at any fixed
finite temperature, $I_{\textrm{topo}}$ is non-zero only when the size of
the system is finite. Next, for fixed finite system size, there is always a
temperature regime where the order parameter assumes a constant value before
dropping to zero. A similar behavior is observed when the size of the system
is increased for a fixed value of the temperature. Importantly, we exhibit a
scaling relation that tells how much the temperature should be decreased to
compensate for an increase of the size of the system, if the system is to
remain topologically ordered. The exact behaviour of topological ordered systems with respect to temperature has received much attention recently. From a {\it static} point of view, the first results indicating the fragility of the 2D toric code against temperature can be found in \cite{nussinov,Cast06}. More recently, the authors of \cite{Kay, Alicki} have shown the fragility of this system during the thermalization {\it dynamics}. In \cite{Kay2, Bravyi} a more general approach is taken to show the impossibility of having a self correcting quantum memory if one restricts the search to 2D stabilizer codes. Apart from providing exact scaling relations, our work is the first to deal with the non-Abelian situation. A different new approach in this respect can be found in \cite{Abasto}. Though $I_{\textrm{topo}}$ is initially introduced as the natural generalization of the topological entropy $S_{\rm topo}$ of Kitaev-Preskill \cite{Kit-Pre05} and Levin-Wen \cite{Levin-Wen05:topent} to the thermal situation, the usefulness of this quantity will become clear throughout the paper. In particular, at least for the toric code, we explicitly relate $I_{\rm topo}$ with the probability distribution of topological sectors in a region, connecting in this way the value of $I_{\rm topo}$ with the ability of a system to perform quantum computing tasks at finite temperature. Moreover, we show how $I_{\textrm{topo}}$ allows to distinguish between quantum double models which share the same value of $S_{\rm topo}$ at zero temperature.

The paper is structured as follows. In Section~\ref{sec:topentropy}, we introduce $I_{\rm topo}$ and discuss some of its properties. In Section~\ref{sec:toric}, we compute $I_{\rm topo}$ for the toric code and show explicitly its dependence on size and temperature. We also explain how to simulate the behavior of $I_{\rm topo}$  in a one-paquette toric code with current technology. In Section~\ref{sec:nonAbelian}, the behavior of $I_{\rm topo}$ for the non-Abelian case is studied. The general formalism is presented and the special case of $D(S_3)$ is given in detail. We provide numerical evidence of the fact that $I_{\rm topo}$ depends on size and temperature exactly as for the toric code. Finally, in Section~\ref{sec:conclusions} we discuss the conclusions and implications of our work.

\section{Topological mutual information}\label{sec:topentropy}

Our analysis is based on the idea that constant corrections to area
laws are typical signatures of topological order. Let us have a closer look
at this property. Consider a bipartition $R:R^c$ of a given system in a pure
state of its ground subspace, and assume that the von Neumann entropy of the
reduced density operator of $R$, $S_R=-\tr \rho_R \ln \rho_R$ satisfies:
\beq\label{eq:arealaw}
S_R=\alpha' |\partial R|-\gamma'+\epsilon(|\partial R|),
\eeq
where $\alpha'$ is a constant, $|\partial R|$ denotes the size of the
boundary of $R$ and where $\epsilon$ tends to zero when $| \partial R|$ tends to infinity. 
As discussed in \cite{Ham04,Levin-Wen05:topent,Kit-Pre05}, systems with a
non-zero value of the topological entropy, $\gamma'$, are topologically
ordered. Indeed, $\gamma'$ is related to the total \emph{quantum dimension}
$\mathcal{D}$ of the anyonic model describing the excitations:
\beq
\gamma'=\ln \mathcal{D}=\ln \sqrt{\sum_{q} d^2_{q}},
\eeq
where $d_q$ is the quantum dimension associated with anyon type $q$
~\cite{Levin-Wen05:topent,Kit-Pre05}. The trivial case with
$\mathcal{D}=1$, i.e. $\gamma'=0$, corresponds to non-topological models,
where the only contribution to $\mathcal{D}$ comes from the vacuum.

\begin{figure}[h]
\begin{center}
\includegraphics[width=30mm,height=30mm]{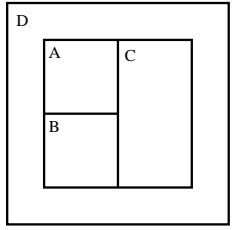}
\caption{Division of a torus or a sphere into four regions.}\label{fig:torusdivided}
\end{center}
\end{figure}

Let us consider a system defined on a closed surface $\Sigma$. In the
remainder of this paper, $\Sigma$ will either be a torus or a sphere. It was
shown in \cite{Levin-Wen05:topent,Kit-Pre05} that $\gamma'$ can be expressed
as a linear combination of entropies of regions of $\Sigma$. For example,
if $\Sigma$ is divided into four regions, as indicated in
Fig.~\ref{fig:torusdivided}, then \cite{Kit-Pre05}
\beq\label{eq:KPgamma}
\gamma' =S_A+S_B+S_C-S_{AB}-S_{AC}-S_{BC}+S_{ABC}.
\eeq
Actually, there is much freedom in constructing linear combinations of entropies that
are topologically invariant, i.e. invariant under local deformations of the
boundaries. It is reasonable to require that the regions $A,B,C$ be treated
on the same footing, and thus write such a linear combination as
$I_{\textrm{topo}}=a^1(S_A+S_B+S_C)+a^2(S_{AB}+S_{AC}+S_{BC})+a^3
S_{ABC}+b^1(S_{AD}+S_{BD}+S_{CD})+b^2(S_{ABD}+S_{ACD}+S_{BCD})+b^3
S_{ABCD}+x S_D$. Consider a deformation of the frontier between regions $C$
and $D$ away from any triple point. Since the deformation is well inside the
region $CD$, we have $\Delta S_{CD}=\Delta S_{ACD}=
\Delta S_{BCD}=\Delta S_{ABCD}=0$. Since it is local between the regions $C$ and $D$, $\Delta S_{C}=\Delta S_{AC}=\Delta S_{BC}=\Delta S_{ABC}$ and
$\Delta S_{D}=\Delta S_{AD}=\Delta S_{BD}=\Delta S_{ABD}$. Therefore
\beq
\Delta I_{\textrm{topo}}= (a^1+2 a^2+ a^3) \Delta S_C+(2 b^1 +b^2+x) \Delta S_D.
\eeq
So, in order to get a topological invariant, we must have that $a^1+2
a^2+a^3=2 b^1+b^2+x=0$. Now let us consider a triple point deformation, at
the intersection between the regions $B,C$ and $D$, say. Reasoning as
before, we get
\bed
\Delta I_{\textrm{topo}}= (a^1+a^2) (\Delta S_B+ \Delta S_C)+
(a^2+a^3) \Delta S_{BC}
\eed
\beq
+ (b^1+b^2) (\Delta S_{BD}+\Delta S_{CD})+(b^1+x) \Delta S_D.
\eeq
We thus get stronger conditions: $a^1=-a^2=a^3$ and $b^1=-b^2=-x$. The
entropy of the total system $S_{ABCD}$ is irrelevant, as expected; it is invariant under boundary deformations. Hence, there is no constraint on
$b^3$. For $a^1=1, b^1=b^3=x=0$, we recover the topological entropy,
$\gamma'$, defined in \cite{Kit-Pre05}. In the following, we work with the
choice $a^1=-b^1=-b^3=x=1$. This choice yields
\beq\label{eq:itopo}
I_{\textrm{topo}}=I_A+I_B+I_C-I_{AB}-I_{AC}-I_{BC}+I_{ABC},
\eeq
which amounts to replace the von Neumann entropies appearing in the
definition of the topological entropy \cite{Kit-Pre05} by quantum  mutual
information. ($I_R$ is defined as $S_R+S_{R_c}-S_{R \cup R_c}$.)

At finite temperature, the von Neumann entropy of a region $R$, $S_R$, is
not a measure of correlations between $R$ and the rest of the system, as it
is in the pure state case. In contrast, the quantum mutual information,
$I_R$, still is. Moreover, $S_R$ does not obey an area law of the form 
(\ref{eq:arealaw}) anymore, whereas for the lattice systems we are going to
study, the mutual information still does \footnote{Note that under quite general assumptions, a
weak form of area law always holds at finite temperature: $I_R \leq \alpha
|\partial R|$, for some constant $\alpha$ \cite{wolf:area}.}. That is, the
properties of the von Neumann entropy that make $\gamma'$ a topological
order parameter at zero temperature are no longer valid at finite temperature.
But they still hold for the quantum mutual information. This is why we
choose to work here with $I_{\textrm{topo}}=\gamma$ instead of $\gamma'$. In Section \ref{sect:topoentropy}, we will further discuss the behaviour of $\gamma'$. 

We close this section by showing that any linear combination of entropies
that is topologically invariant should, in general, involve a division of
the surface $\Sigma$ into at least \emph{four} regions. It will be enough to
consider the case where the whole system is in a pure state. We would like
to construct a particular linear combination of entropies that isolates
$\gamma$ from the area part. It is clear that partitioning the surface into
a region $R$ and its complement $R^c$ cannot provide such a quantity.
Indeed, in this case $S_R=S_{R_c}$, $S_{R \cup R_c}=0$ and $\partial R$ and
$|\gamma|$ have a common fate. Three regions $A$, $B$ and $C$ are not
sufficient neither. This is easily shown considering a system in a pure
state. Let $l_A, l_B$ and $l_{AB}$ denote respectively the length of the
boundary between region $A$ and region $C$, region $B$ and region $C$, and
region $A$ and region $B$, and let $n_A$ denote the number of connected
pieces that make region $A$, $n_B$ and $n_{AB}$ are defined likewise.
From the relations $S_A+S_B=\alpha(l_A+l_B+l_{AB})-(n_A+n_B) \gamma$,
$S_A-S_B=\alpha(l_A-l_B)+(n_A-n_B) \gamma$ and
$S_{AB}=S_C=\alpha(l_A+l_B)-n_{AB} \gamma$, we see that it is impossible to
construct a linear combination of entropies that cancels all boundary
contributions and leaves only the topological contributions. Thus, four is
the minimal number of pieces required in order to partition $\Sigma$ in such
a way that it gives a boundary-independent quantity in a non-trivial way.

\section{The toric code}\label{sec:toric}

The toric code is a simple topological model, with a Hamiltonian that can be
diagonalized exactly \cite{Kit97}. It is called a `code' because it is a
quantum error correcting code; two logical qubits are encoded in the
physical system. An optical implementation of a four-qubit toric code is
possible since its ground state is a GHZ state
\cite{oneplaquette}, and proposals to create the ground state of large toric
codes, as well as elementary excitations in optical lattices are described
in \cite{aguado:imp}. In this section, we study how $I_{\textrm{topo}}$
behaves as the (inverse) temperature, $\beta$, and the size of the system,
$L$, are varied. A similar calculation has been presented in \cite{Cast06}.
We nevertheless present our alternative approach in details because it
differs in two important respects. First, it is $\gamma'$ which is used in
\cite{Cast06} as a topological order parameter. Second, the simplicity of
our alternative analysis has allowed us to get analytically more general as
well as new results such as the scaling laws discussed in Section \ref{subsec:scaling}. 
Moreover, this calculation helps understand better how to compute the topological 
mutual information for non-Abelian models.

\subsection{Spectrum}

\begin{figure}[h]
\begin{center}
\includegraphics[width=80mm,height=30mm]{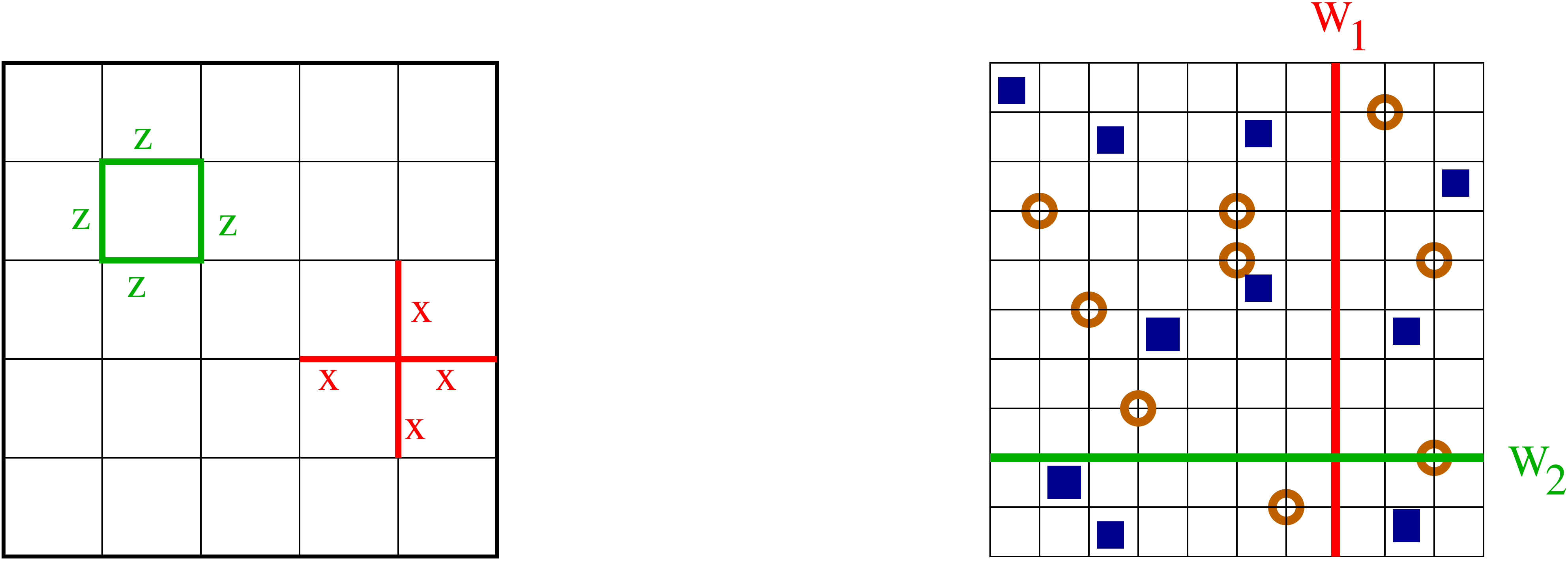}
\end{center}
\caption{(Left) Pictural representation of the $B_p$ and the $A_s$ operators. (Right) An eigenstate of the toric code lattice. Elementary excitations can be either of flux type, depicted by squares in plaquettes, or of charge type depicted by hollow dots on vertices. The strings along the torus correspond
to Wilson loops.}\label{fig:gasdefect}
\end{figure}

Let us consider a torus tiled into $L \times L$ square plaquettes and
associate a two-level system (qubit) with each edge of the obtained lattice.
We assume that these qubits interact through the Hamiltonian
\beq\label{eq:toric}
H=-J\sum_{p} B_p- J' \sum_{s} A_s,
\eeq
where the index $p$ (resp. $s$) runs over all plaquettes (resp. vertices) of
the tiling. The operator $B_p$ involves all the spins surrounding the
plaquette $p$, while the operator $A_s$ involves all the spins with one end
at $s$. They are defined as follows:
\beq
B_p=\prod_{i \in p} \sigma^z_i,
\hspace{1cm}
A_s=\prod_{i \in s} \sigma^x_i,
\eeq
and represented on Fig.\ref{fig:gasdefect}. The coupling constants $J$ and $J'$ will be chosen to be both equal to 1, for the sake of simplicity. But all our analysis can be straightforwardly generalised to arbitrary values of $J$ and $J'$.

Due to the topology of the torus, we have that
\beq\label{eq:consttorus}
\Pi_p B_p= \Pi_s A_s=1.
\eeq
This constraint means that the number of excited plaquettes (resp. excited
vertices) is always even. The Hamiltonian $H$ is a sum of local terms all
commuting with each other.  We observe that although this property
facilitates the diagonalization of $H$, it is not sufficient to guarantee
that it is easy to solve. What makes $H$ exactly diagonalizable is that, as
it turns out, its eigenstates of $H$ can have arbitrary eigenvalues of the
operators $B_p$ and $A_s$, up to the constraint (\ref{eq:consttorus}).
Therefore, each eigenstate of $H$ is given by a triple of ``quantum
numbers'': $\ket{\phi,c, w}$, see Fig.~\ref{fig:gasdefect}. A pattern $\phi$
denotes the position of all plaquette or ``flux-type'' excitations. Another
pattern, $c$, indicates the position of all vertex or ``charge-type''
excitations. Finally, $w$ indexes the degeneracy of the state for a fixed
configuration of defects. This quantum number is made of two bits, $w_1$ and
$w_2$, that label the values of the integrals of motion of $z$-operators
around non-contractible loops on the torus \cite{Kit97} (Wilson loops). We
have
\beq
H \ket{\phi,c,w}= (E_0+2 |\phi|+2 |c|) \; \ket{\phi,c,w},
\eeq
where $E_0=-2L^2$ is the ground state energy and $|\phi|$ (resp. $|c|$)
denotes the number of flux excitations (resp. charge excitations) of the
pattern $\phi$ (resp. $c$). The eigenvalues of $H$ satisfy $E_n-E_{n+1}=-4$
and range between $-2 L^2$ and $2 L^2$. Also, if $P_i$ denotes the projector
onto the sector of energy $E_i=E_0+4 i$ and $d_i=\tr P_i$ denotes its
dimension, we have that
\beq
d_i=4 \sum_{n_{\phi},n_{c} \leq L^2/2} \; \sum_{n_{\phi}+n_{c}=i} \binom{L^2}{2 n_{\phi}}
\binom{L^2}{2 n_{c}}.
\eeq
One can check that $\sum_i d_i=2^{2 L^2}$.

\subsection{von Neumann entropy}

In the following, we shall consider a situation in which the system is immersed in a bath at inverse temperature $\beta$ and is let to thermalize. Since $H=\sum_{i=0}^{L^2} E_i P_i$, the partition function of this model reads
\beq\label{eq:defzz2}
Z(\beta,L)=\tr \; e^{-\beta H}=\sum_{i=0}^{L^2} e^{-\beta E_i} d_i,
\eeq
This series can be easily summed up (see \cite{nussinov} or details in Appendix \ref{sec:detailstoric}). We get
\beq
Z(\beta,L)=((2 \cosh \beta)^{L^2}+(2 \sinh \beta)^{L^2})^2.
\eeq
The thermal state of Eq.(\ref{eq:toric}) reads
\beq\label{eq:thermalstatetoric}
\rho_{\textrm{th}}= e^{-\beta H}/Z(\beta,L).
\eeq
The von Neumann entropy, $S_{\textrm{tot}}=-\tr \rho_{\textrm{th}} \ln \rho_{\textrm{th}}$, of the whole torus is then easily derived from the partition function thanks to the identity
\beq\label{eq:Zentropy}
S_{\textrm{tot}}=-\frac{\beta}{Z(\beta,L)} \frac{\partial}{\partial \beta} Z(\beta,L)+\ln Z(\beta,L).
\eeq

\begin{figure}[h]
\begin{center}
\includegraphics[width=30mm,height=15mm]{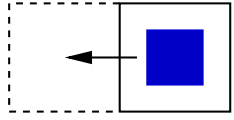}
\end{center}
\caption{Solid edges belong to a region $R$, while dashed edges belong
to $R_c=\Sigma \backslash R$. A plaquette  excitation on the right $\partial
R$ plaquette can be driven inside $R_c$ applying $\sigma^x$ on the edge
separating $R$ from $R_c$.}\label{fig:excitab}
\end{figure}

We now compute the von Neumann entropy of a connected region $R \subset \Sigma$, $S_R$. A couple of observations about the reduced state, $\rho_R$, allows to get an analytic expression for $S_R$. First, we consider a fixed eigenstate $\ket{\phi,c,w}$ of the Hamiltonian (\ref{eq:toric}). The reduced state $\rho_R(\phi,c,w)=\tr_{R_c} \ket{\phi,c,w}
\bra{\phi,c,w}$ does not depend on $w$ if $R$ is contractible; homologically
non-trivial loops are necessary to measure $w$. Also, two states $(\phi,c,w)
\neq (\phi',c',w')$ are orthogonal whenever $(\phi_R,c_R) \neq
(\phi'_R,c'_R)$ since they can be discriminated by measuring $A_s$ operators
or $B_p$ operators having support on $R$. Next, it is useful to distinguish
three kinds of plaquette excitations: those with support fully on
$R$, $\phi_R$, those with support fully on $R_c$, $\phi_{R_c}$, and the
others, $\phi_{\partial R}$. Similarly, we divide vertex excitations into
three  kinds: $c_R, c_{R_c}, c_{\partial R}$. In order to lighten the
notations, we use the symbol $\gras{q}$ to label configurations of defects,
both plaquette and vertex, i.e. $\gras{q} \equiv (\phi,c)$. Crucially,
excitations of the $\partial R$ type can be driven inside $R_c$ by
application of Pauli operators acting on links of $R_c$, that is:
\beq
\ket{\gras{q}_R,\gras{q}_{R_c},\gras{q}_{\partial R},w}= U''_{R_c} \ket{\gras{q}_R, \gras{q}'_{R_c},w},
\eeq
for some unitary operator $U''_{R_c}$ and some configuration of defects in
$R_c$, $\gras{q}'_{R_c}$. This property is illustrated in
Fig.~\ref{fig:excitab}.

All excitations inside $R$ can be fused into a single excitation $q_R^1$
located on a site\footnote{As in \cite{Kit97}, we call a site a combination
of a vertex and an adjacent plaquette.}, using a unitary whose support is
fully in $R$ and similarly for $R_c$. In turn, the state
$\ket{q_R^1,q_{R_c}^1,w}$ can be created from a ground state $\ket{\xi,w}$
by application of Pauli operators along strings connecting the site where
$q_R^1$ is located to the site where $q_{R_c}^1$ is located. So,
\beq
\ket{\gras{q},w}=U_R(\gras{q}_R) \otimes U_{R_c}(\gras{q}_{R_c},\gras{q}_{ \partial R}) \ket{\xi,w}
\eeq
for some unitary operators $U_R(\gras{q}_R)$ and $U_{R_c}(\gras{q}_{R_c},\gras{q}_{ \partial R})$. We are now in a position to characterise $\rho_R$ and compute $S_R$. The thermal state of the toric code can be written explicitly as
\beq
\rho_{\textrm{th}}=\sum_{w,\gras{q}} \frac{e^{-\beta(E_0+ \Delta E |\gras{q}|)}}{Z(\beta,L)}
\ket{\gras{q},w}\bra{\gras{q},w},
\eeq
where $\Delta E=2$ is the energy associated with a single excitation (plaquette or vertex). Therefore, for a non-contractible region $R$,
\beq\label{eq:rhoRdecompo}
\rho_R= \sum_{w,\gras{q}_R} C(q_R) \tr_{R_c} [U_R(\gras{q}_R)  \ket{\xi,w}\bra{\xi,w}
U_R(\gras{q}_R)^{\dagger}],
\eeq
where 
\bed
C(\gras{q}_R)=\sum_{\gras{q}_{R_c},q_{\partial R}} e^{-\beta(E_0+2
|\gras{q}_R|+2 |\gras{q}_{R_c}|+2 |\gras{q}_{\partial R}|)}/Z(\beta,L).
\eed
Note that $4 C(\gras{q}_R)$ is the marginal probability of a configuration
of defects $\gras{q}_R$. The decomposition (\ref{eq:rhoRdecompo}) allows to
compute $S_R$. Indeed, from the identity
\beq\label{eq:entropydirectsum}
S \big( \bigoplus_i \lambda_i \rho_i \big)=-\sum_i \lambda_i \ln \lambda_i+\sum_i \lambda_i S(\rho_i),
\eeq
we find that
\beq\label{eq:entropyR}
S_R=S_R^{\textrm{gs}}-\sum_{\gras{q}_R} 4 C(\gras{q}_R) \ln(4 C(\gras{q}_R)),
\eeq
where $S_R^{\textrm{gs}}$ is the von Neumann entropy of the region $R$ when
the system is in a pure ground state $\ket{\xi,w}$:
$S_R^{\textrm{gs}}=(|\partial R|-1) \ln 2$ \cite{Ham04}.

It turns out that the sums appearing in Eq.(\ref{eq:entropyR}) can be carried out exactly (see details in Appendix \ref{sec:detailstoric}). The result is that the entropy of a region can be expressed as
\beq\label{eq:entropyfinitet}
S_R=S_R^{\textrm{gs}}+V(\beta,N_p(R),L)+V(\beta,N_*(R),L),
\eeq
with
\bed
V(\beta,N,L)= N \ln(1+e^{-2 \beta})+ \ln(1+\theta^{L^2})
+\frac{N \beta e^{-\beta}}{\cosh\beta} \frac{1-\theta^{L^2-1}}{1+\theta^{L^2}}+ \ln2
\eed
\bed
-\frac{(1+\theta^N) (1+\theta^{L^2-N})}{2 (1+\theta^{L^2})}
\ln(1+\theta^{L^2-N})
\eed
\beq\label{eq:entropytorus}
-\frac{(1-\theta^N) (1-\theta^{L^2-N})}{2 (1+\theta^{L^2})}
\ln(1-\theta^{L^2-N}),
\eeq
where $\theta=\tanh\beta$.

As we can see, the entropy of a region separates neatly into a pure state
contribution and a finite temperature contribution. The first only involves
the area of the region, while the second depends on its volume ($N_p(R)$ and
$N_*(R)$). By increasing the temperature, we pass from an area law to a volume law, as expected. The calculations are almost identical when $R$ is only semi-contractible. The only difference is that $\rho_R$ will depend on only one of the Wilson loops, and the entanglement entropy picks a $-\ln 2$ additive correction. When $R$ is completely contractible, the correction is $-\ln 4$.

\subsection{Dependence of $I_{\textrm{topo}}$ on size and temperature}\label{subsec:scaling}

\begin{figure}[h]
\begin{center}
\includegraphics[width=60mm,height=48mm]{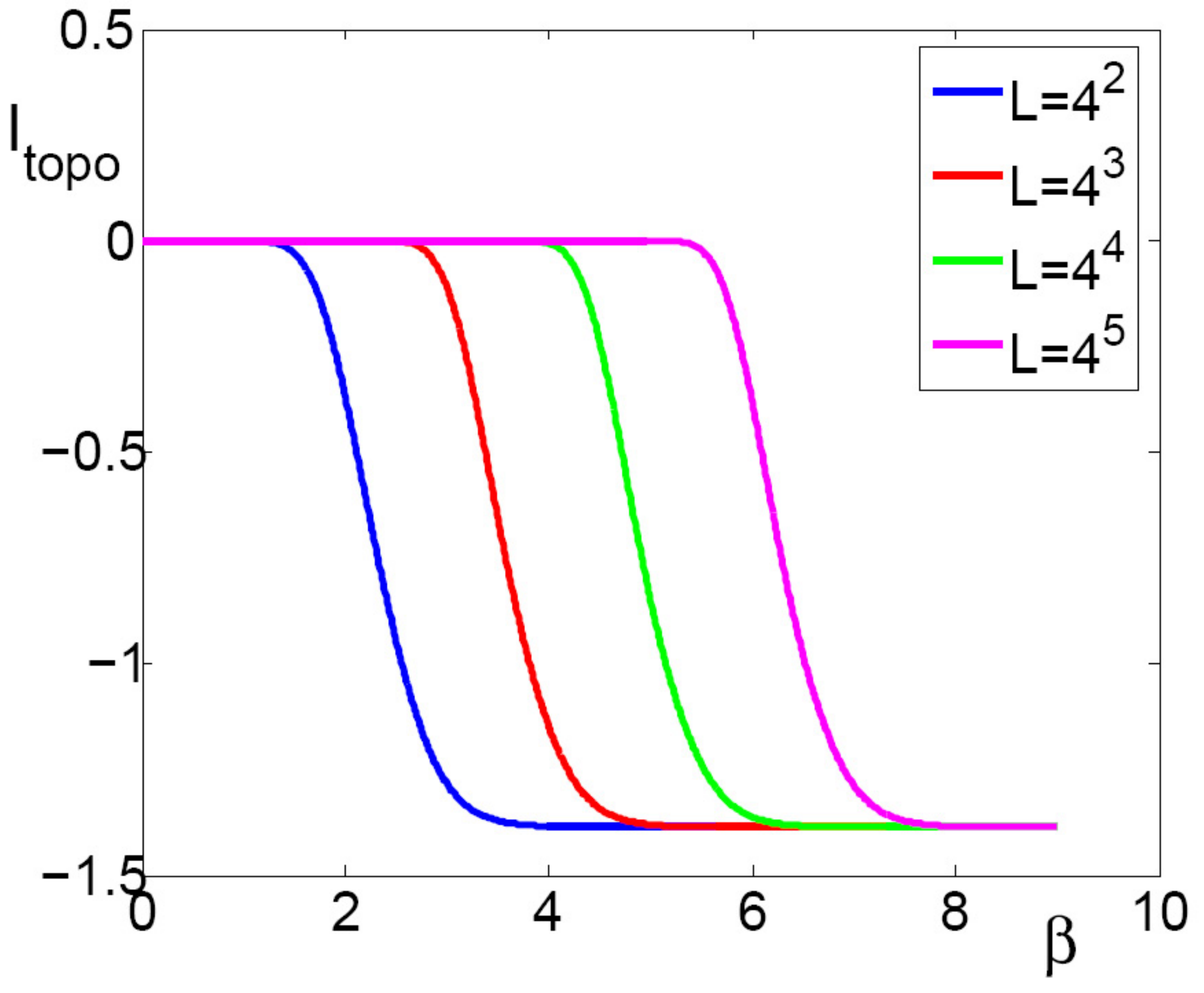}
\caption{Topological mutual information as a function of $\beta$ for tori
of size $4^2 \times 4^2$, $4^3 \times 4^3$, $4^4 \times 4^4$ and $4^5 \times
4^5$. The size of the region A, as indicated in Fig.~\ref{fig:torusdivided}
is $k \times k= (L/4) \times (L/4)$.}
\label{fig:puretopentropy}
\end{center}
\end{figure}

Using Eq.(\ref{eq:entropytorus}), we have plotted the topological mutual
information (\ref{eq:itopo}) for tori of various sizes, see
Fig.~\ref{fig:puretopentropy}. For a fixed torus size, we observe that there
is a region of values of $\beta$ such that $I_{\textrm{topo}}$ is non-zero
and stationary. Then, as the temperature is increased, $I_{\textrm{topo}}$
smoothly vanishes. Interestingly, the transition does not become more abrupt
when the size of the system is increased. Rather, the curves displayed are similar.

From Fig.~\ref{fig:puretopentropy}, we also see that, when the system size
increases, the temperature at which the transition occurs decreases.
Actually, in the limit of large codes, the transition temperature vanishes.
When systems described by a local hamiltonian are left in thermal
equilibrium, their mutual information is bounded by a constant times their
area \cite{wolf:area}. As it turns out, the systems we are studying obey a
strict area law, i. e. the quantum mutual information between a region $R$
and the rest of the sytem satisfies:
\beq\label{eq:arealawqinfo}
I_R=\alpha(\beta,L) |\partial R|- \gamma(\beta,R,L).
\eeq
For the toric code, we can compute
\bed
\alpha_{\infty}(\beta)= \lim_{|\partial R| \to \infty} I_R/|\partial R|,
\hspace{1cm}
\gamma_{\infty}(\beta)=\lim_{|\partial R| \to \infty} (I_R-\alpha_{\infty}(\beta) |\partial R|).
\eed

Interestingly, we find slightly different results for $\gamma_{\infty}(\beta)$, depending on how this limit is taken. One possibility is to first consider the limit for $L \to \infty$, and then let the size of the region $R$ grow. One finds:
\bed
\gamma(\beta,R,\infty)=2 \ln2
\eed
\bed
+\frac{1+\theta^{\bar{N}_*(R)}}{2} \ln \frac{1+\theta^{\bar{N}_*(R)}}{2}
+\frac{1-\theta^{\bar{N}_*(R)}}{2} \ln \frac{1-\theta^{\bar{N}_*(R)}}{2}
\eed
\beq\label{eq:asymp1}
+\frac{1+\theta^{\bar{N}_p(R)}}{2} \ln \frac{1+\theta^{\bar{N}_p(R)}}{2}
+\frac{1-\theta^{\bar{N}_p(R)}}{2} \ln \frac{1-\theta^{\bar{N}_p(R)}}{2},
\eeq
where $\theta=\tanh \beta$, and where $\bar{N}_p(R)=L^2-N_p(R_c)$ denotes the number of plaquettes within a region $R$ and at its border. $\bar{N}_*(R)$ is defined likewise.

Eq. (\ref{eq:asymp1}) lends itself to a simple interpretation. Let $p_e$ denote the probability that a plaquette or a site is excited.  The mean energy of the system reads $\mean{H}=E_0+4 p_e L^2$. From $\mean{H} Z(\beta,L) =-\partial Z(\beta,L)/\partial \beta$, we find that in the limit $L \to \infty$, $p_e=(1-\theta)/2$. On another hand, ignoring total anyonic charge conservation, the probability that a region $R$ containing $N_p(R)$ plaquettes has an even number of excited plaquettes reads
\beq
p^{\textrm{even}}_{p}(R)=\frac{1}{2} \sum_{i=0}^{N_p(R)} \binom{N_p(R)}{i} (p_e^i+(-p_e)^i)
(1-p_e)^{N_p(R)-i}=\frac{1}{2}(1+\theta^{N_p(R)}).
\eeq
One can similarly calculate $p^{\textrm{even}}_*(R)$, the probability that the region $R$ contains an even number of excited vertices. One find the same expression with $N_p(R)$ replaced by $N_*(R)$. So the asymptotic limit of the topological mutual information can be rewritten as
\beq\label{eq:simpletopo}
\gamma_{\infty}(\beta)=2 \ln2-h_2(p_{p}^{\textrm{even}}(R))-h_2(p_{*}^{\textrm{even}}(R)),
\eeq
where $h_2(x)=-x \ln x-(1-x) \ln (1-x)$ is the Shannon entropy of a binary outcome probability distribution, or equally simply as
\beq\label{eq:simpletopo2}
\gamma_{\infty}(\beta)=-D(\{ p_R(q) \} || \{ p_u(q) \}),
\eeq
where
$\{ p_u \}$ is the four-event uniform probability distribution, $\{ p_R \}$ denotes the thermal probability distribution associated with all possible values for the total anyonic charge: total plaquette flux in $R$ trivial and total electric charge in $R$ trivial, etc, and where $D(\{ p^1 \} || \{ p^2 \})=-\sum_j p^1_j \ln(p^1_j/p^2_j)$ denotes the Kullback-Leitner pseu\-do-di\-sta\-nce between two probability distributions $\{ p^1 \}$ and $\{ p^2 \}$ \cite{Cover}. We can see that $\lim_{k \to \infty} \gamma_{\infty}(\beta)=0$. We conjecture that the formula (\ref{eq:simpletopo2}) is valid in general, with the probability distribution induced by the quantum dimensions:
\beq\label{eq:conjecture}
p_u(q)=\frac{d^2_q}{\mathcal{D}^2}.
\eeq
This relation has been verified for all Abelian quantum double models.

The other possibility, when studying the asymptotic behavior of the
topological mutual information, is to let the size of the torus and the size
of the region $R$, grow at the same rate. Let $\nu L \times \nu L$ denote
the area of the region $R$ ($\nu<1$). In that case, keeping $\theta^{L^2}$
fixed is the only way to make the limit meaningful. Using the fact that
$\lim_{L
\to \infty} \theta^L=1$ in that case, one finds again
Eq.(\ref{eq:simpletopo2}), but the probabilities are slightly different now.
For example
\beq\label{eq:probaasym2}
p^{*}_{\textrm{even}}(R) \simeq \frac{(1+ \theta^{\nu^2 L^2})(1+ \theta^{(1-\nu^2) L^2})}{2(1+\theta^{L^2})},
\hspace{0.5cm}
p^{p}_{\textrm{even}}(R) \simeq p^{*}_{\textrm{even}}(R).
\eeq
These quantities are still the probabilities corresponding to the value of
the total charge (flux) sector for region $R$ but subject to the global flux
(charge) neutrality condition.

Eqs. (\ref{eq:simpletopo2},\ref{eq:probaasym2}) are very interesting in that
they allow to extract a \emph{scaling law} for the topological mutual
information. In the simultaneous limit, i.e. for a fixed value of $\nu$, the
topological mutual information only depends on the temperature and size
through the parameter $t=\tanh(\beta)^{L^2}$. In particular, a fixed value
of $t$, and thus a fixed value of the topological mutual information,
corresponds to the following relation between size and temperature
\bed
\beta(t,L)=\ln L-\frac{1}{2} \ln(\frac{1}{2} \ln \frac{1}{t})+O(L^{-2}),
\eed
\beq\label{eq:scaling}
\frac{\partial T(t,L)}{\partial L}=\frac{-1}{L (\ln L-\frac{1}{2} \ln(\frac{1}{2} \ln \frac{1}{t})+O(L^{-2}))^2}+O(L^{-2}).
\eeq

For a general Abelian quantum double, based on a group $G = \mathbb{Z}_{K_1} \times \ldots \times \mathbb{Z}_{K_r}$, these formulae generalise to 
\bed
\beta(t,L)=\ln L-\frac{1}{2} \ln(\frac{1}{K} \ln \frac{1}{t})+O(L^{-2}),
\eed
where $K=K_1 \ldots K_r$, and where the scaling variable is now defined as 
\bed
t=\big( \frac{1-e^{-\beta \Delta E}}{1+(K-1) e^{- \beta \Delta E}} \big)^{L^2},
\eed
where $\Delta E$ denotes again the energy associated with an excitation.

These relations tell us how an increase of the size of the system should be
compensated by a decrease of temperature in order to maintain a fixed value
of the topological mutual information. We believe that they constitute a qualitative
nuance from the fact that at any fixed finite temperature, the topological
mutual information, or the topological entropy, asymptotically vanishes when the
size of the system is increased \cite{Cast06}. In particular, they show that
the \emph{rate} at which the temperature should be decreased, in order to
maintain a fixed value of $\gamma$, \emph{decreases} with the size of the
system.

\subsection{Mean value of the $S$-matrix }

As we have seen, the sub-leading term in the area law for $I_{R : R^c}$
contains entropic information about the thermal probability distribution of
topological sectors inside region $R$.  In fact, we now wish to argue that
this probability distribution is the fundamental quantity controlling in
principle (up to implementation-specific problems) the usefulness of
topological quantum memory and quantum computing by anyon braiding.

\begin{figure}[h]
\centering
\includegraphics[scale=.3]{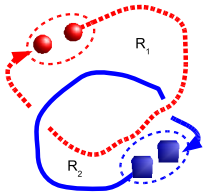}
\caption{\label{fig:winding} Braiding of anyons in the toric code.  A
  pair of electric defects (spheres) and a pair of magnetic defects
  (cubes) are created.  One electric and one magnetic defect are wound
  around each other, then both pairs are annihilated.  The anyon
  trajectories define a two-dimensional projection of a Hopf link,
  whose components enclose regions $R_1$ and $R_2$.}
\end{figure}

For concreteness we take the toric code model that has been initially
proposed as topological memory~\cite{Kit97}. Consider the process of
creating a pair of electric and a pair of magnetic defects, braiding between
one particle of each pair and subsequently annihilating both pairs, as shown
in Fig.~\ref{fig:winding}. The crucial property of these anyons is that the
operation $U_{\mathrm{Hopf}}$ effecting this process on any state of the
ground level multiplies it by $-1$; indeed, $U_{\mathrm{Hopf}} = -
U^{\mathrm{e}}_{R_1} U^{\mathrm{m}}_{R_2}$, where $U^{\mathrm{e}}_{R_1}$ and
$U^{\mathrm{m}}_{R_2}$ are the operators describing the processes associated
with the electric and magnetic anyons separately, i.e., unlinked.
Thus, the thermal expectation values are immediately related to (marginals
of) the charge probability distributions since $U^{\mathrm{e}}_{R_1}$ and
$U^{\mathrm{m}}_{R_2}$ measure the electric and magnetic charges inside
their regions:
\begin{equation}\label{eq:thermalwindings}
  \langle U_{\mathrm{Hopf}} \rangle_\beta
=
 - \,
  \langle U^{\mathrm{e}}_{R_1} \rangle_\beta
  \langle U^{\mathrm{m}}_{R_2} \rangle_\beta
=
 - \,
  ( p^*_{\textrm{even}}(R_1) -  p^*_{\textrm{odd}}(R_1))
  ( p^p_{\textrm{even}}(R_2) -  p^p_{\textrm{odd}}(R_2))
 \; .
\end{equation}

As a consequence, the expectation value of $U_{\mathrm{Hopf}}$ is controlled
by scaling variables $\theta^{\mathrm{vol}(R_1)}$ and
$\theta^{\mathrm{vol}(R_2)}$.  We expect this will be the case for many anyonic models, 
where similar expectation values control the \emph{visibility} of
interferometry experiments \cite{interferometry}. Note that Hopf-link-like processes
define the elements of the topological $S$-matrix and twisted self-braiding
of anyons yields their topological spin.  Thus, these fundamental quantities
of the anyon model are degraded at finite temperature at a rate controlled
by the thermal charge probability distributions. The latter becomes the
object that determines the appropriateness of the system to perform quantum
computation at finite temperature.

In particular, $I^{\mathrm{topo}}$ measures the Shannon entropy of this
distribution. This is why we believe that it is a good topological order
parameter. Note that in the ground level, the probability of finding anyons
in a given region vanishes, therefore the probability of a given sector $q$,
$p^q(R)$, becomes $\delta_{q, 1}$ and  the distribution has zero Shannon
entropy. For high temperatures the distribution approaches
(\ref{eq:conjecture}), which is  dictated only by the quantum dimensions of
the anyons. Our conjecture (\ref{eq:simpletopo2}) implies that this is the
maximal Shannon entropy available to the distribution.

\subsection{Topological Entropy}\label{sect:topoentropy}

We now analyze the behavior of $\gamma'$ introduced in
Eq.(\ref{eq:KPgamma}). We start by observing that if one writes $\gamma'$ as
$\sum_R \sigma_R S_R$, where $R \in
\{  A,B,C,AB,AC,BC,ABC \}$ (see Fig.~\ref{fig:torusdivided}), then
\beq
\sum_R \sigma_R (N_p(R)+N_*(R))=1.
\eeq
With this lattice relation and Eq.(\ref{eq:entropyfinitet}), one can see that in the
limit where the sizes of all regions diverge,
\beq
\gamma' \to \ln 2 -\frac{2 \beta}{e^{2 \beta}+1}-\ln(1+e^{-2 \beta}).
\eeq
At fixed temperature, $\gamma'$ does \emph{not} vanish as the size of the
system grows, a behavior that contrasts with that of $I_{\textrm{topo}}$. It actually becomes independent of the system size. Since, at finite temperature, the von Neumann entropy of a region is no longer a measure of its correlations with the rest of the system, it is not clear whether $\gamma'$ actually still probes topological order. The discrepancy between the behaviour of $\gamma$ and that of $\gamma'$ is made obvious here because magnetic and electric defects have been treated on an equal footing  \emph{right from the start} ($J=J'$ in Eq.(\ref{eq:toric})). Our results therefore do not contradict those of \cite{Cast06}). Note that $\gamma'$ vanishes in the limit where $\beta$ tends to zero, as expected.

\subsection{The Case of One Plaquette}

In this subsection we develop a quantum simulation of a minimal toric code at finite temperature \cite{oneplaquette}. We therefore consider a four qubit GHZ state, coupled to a single ancillary qubit, that plays the role of the environment. For a particular time-dependent coupling between the system and the environment, it is possible to reproduce the exact behavior of topological entanglement as a function of temperature, where the latter is now represented by time. This can be also viewed as a purification protocol of the topological thermal states. An experimental verification of this topological behavior could be demonstrated with state-of-the-art technology.

\begin{figure}[h]
\begin{center}
\includegraphics[width=8cm,height=4cm]{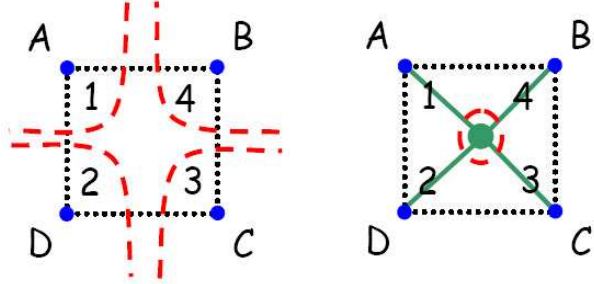}
\caption{(Left) The one plaquette system of the toric code, which is in
the GHZ state. Each subsystem $A$, $B$, $C$ and $D$ consists in a single
qubit. (Right) The same system without the Hamiltonian~(\ref{Ham1}) coupled
to the environment, here taken to be one extra ancillary qubit. The latter
can be coupled to all the plaquette qubits or equivalently to just one of
them.}
\label{plaquette}
\end{center}
\end{figure}

Consider a single plaquette of the toric code model \cite{oneplaquette}. The corresponding
Hamiltonian can be given by
\be
H=-J\sigma^x_1\sigma^x_2\sigma^x_3\sigma^x_4-\sigma^z_1\sigma^z_2-
\sigma^z_2\sigma^z_3-\sigma^z_3\sigma^z_4
\label{Ham1}
\ee
where we have omitted the term $\sigma^z_4\sigma^z_1$ as it is superfluous
for generating a topologically ordered ground state. The corresponding ground
state (vacuum) is the GHZ state
\be
\ket{\xi}=|GHZ_4\rangle={1 \over \sqrt{2}} (\ket{0000}+\ket{1111}).
\label{GHZ}
\ee
When the plaquette is occupied by an electric charge, the state of the system is
$\ket{e}=\sigma^z_1\ket{\xi}={1 \over \sqrt{2}} (\ket{0000}-\ket{1111})$. It
is possible to consider the thermal state of the system at temperature
$T=1/\beta$,
\be
\rho ={e^{-H\beta} \over \text{Tr}(e^{-H\beta})}.
\ee
We can calculate the topological mutual information for this system.
Consider the splitting of the four qubits in the regions $A$, $B$, $C$ and
$D$, as seen in Fig.~\ref{plaquette}. 

\begin{figure}[h]
\begin{center}
\includegraphics[width=5cm,height=3.5cm]{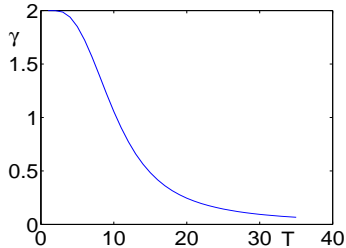}
\caption{Topological mutual information, of the plaquette system as a function of the temperature, $T$. When the Hamiltonian~(\ref{Ham1}) is present we observe a resilience of the
topological character ($\gamma\sim2$) for small temperatures, which is lost
for higher temperatures ($\gamma\sim0$).}
\label{temperature}
\end{center}
\end{figure}

If we plot $I_{\textrm{topo}}$ as a function of the temperature for this system, we get Fig.\ref{temperature}.

Since all the interaction  terms in the Hamiltonian~(\ref{Ham1}) commute with each
other it is easy to evaluate the density matrix. For simplicity we focus on
the case where $J\ll1$ which reduces the density matrix to
\begin{equation}
\rho_\beta= {1\over 2} (1\!\!1
+\sigma^x_1\sigma^x_2\sigma^x_3\sigma^x_4\tanh \beta).
\end{equation}

We would like to simulate the same behavior, but without the background
Hamiltonian~(\ref{Ham1}). We can reproduce the density matrix, $\rho_\beta$,
with four non-interacting qubits initially in state $\ket{\xi}$ coupled to
an environment. We take the environment to be an ancilla initially in state
$\ket{\psi_a}=(\ket{0}+\ket{1})/\sqrt{2}$ coupled to the system through the
interaction
\be
H_\text{int}= \omega \sigma^z_1\sigma^z_a.
\label{Ham2}
\ee
Note that an interaction Hamiltonian that couples the ancilla symmetrically
with all the qubits would give the same results. For
$U(t)=\exp(-i H_\text{int} t)$ the time evolution due to~(\ref{Ham2}) gives a
reduced density matrix for the four qubits of the form
\be
\rho_t=\tr_a(U(t)\ket{\xi,\psi_a}\bra{\xi,\psi_a}U^\dagger(t)).
\ee
For particular choices of a time dependent $\omega$, it is possible to make
$\rho_t$ identical to $\rho_\beta$. Now, the time evolution of $\rho_t$ is
identified with the increase in the temperature, $T$, of $\rho_\beta$. Indeed, if we choose the coupling as
\be
\omega(t) = {J \over t^2}\cosh^{-2} {J \over t},
\ee
then the temporal evolution of the system without the
Hamiltonian~(\ref{Ham1}) is equivalent to constantly increasing the
temperature of the same system in the presence of the Hamiltonian.

Thus, the toric code plaquette with Hamiltonian~(\ref{Ham1}) in a thermal
state at finite temperature, $T$, has the same topological mutual information as a
plaquette initially prepared in $\ket \xi$ and coupled to an ancilla in state $\ket +$, where time plays the role of temperature. By employing larger states and more ancillae, one could reproduce the finite temperature topological behavior of larger toric code systems.

\section{Non-Abelian models}\label{sec:nonAbelian}

We now turn to a class of models featuring non-Abelian anyonic statistics
\cite{Kit97}. We call these models ``non-Abelian superconductors", as in
\cite{Pres:chap9} because of the braiding properties of their excitations.
We mainly focus on a particular model based on the quantum double $D(S_3)$, because it is the simplest in the family we are considering. This model is paradigmatic though; our analysis
can be straightforwardly generalized to models defined through the quantum
double of any finite group. The $D(S_3)$ model is also important from a
quantum information perspective because it allows to perform universal
quantum computation \cite{mochon}. We start by briefly reviewing some properties of the hamiltonians we are considering. Then we show how the entropy of a region can be calculated. This result is
employed to study the behavior of the topological mutual information at
finite temperature.

\subsection{The $D(S_3)$ superconductor and its eight sectors}

\begin{figure}[h]
\begin{center}
\includegraphics[width=28mm,height=32mm]{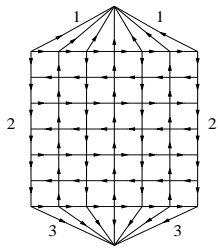}
\end{center}
\caption{Tiled sphere. Equal numbers refer to identified sets of edges.}\label{fig:sphere}
\end{figure}

Consider an oriented lattice, $\Lambda$, on a sphere such as the one
represented in Fig.~\ref{fig:sphere}. Let $G$ denote a finite group, where
$e$ refers to its neutral element. With each edge of $\Lambda$, we associate
a $|G|$-dimensional Hilbert space, $\mathscr{H}$, with an orthonormal
`computational' basis labelled by the group elements $\{\ket{g}: g \in G\}$.
We now borrow a series of definitions introduced in
\cite{Kit97}. We introduce the operators $L_{\pm}$ and $T_{\pm}$, which act
on the computational basis as $L_+(g) \ket{g'}=  \ket{g g'}$, $T_+(g)
\ket{g'}=
\delta_{g g'} \ket{g'}$, $ L_-(g) \ket{g'}=  \ket{g' g^{-1}}$, $T_-(g)
\ket{g'}= \delta_{g^{-1} g'} \ket{g'}$. We also define $L^g(j,s)= L_-(g)$ if
$s$ is the origin of an edge $j$, whereas $L^g(j,s)= L_+(g)$ if $s$ is the
endpoint of $j$. Also, if a plaquette $p$ is at the left (resp. right) of an
edge $j$, we define $T_g(j,p)= T_-(g)$ (resp. $T_g(j,p)= T_+(g)$). We
further define
\beq
A_g(s)=\bigotimes_{j \in *(s)} L^g(j,s), \hspace{0.5cm}
B_g(s,p)=\sum_{g_1 \ldots g_m=g} \; \bigotimes_{m \in \square(p)} T_{g_m}(j_m,p),
\eeq
where in the definition of $B_g(s,p)$, $j_1, \ldots, j_m$ are the boundary
edges of $p$ listed in counterclockwise order, starting from and ending at
some vertex $s$. The operators $A_g(s)$ and $B_g(s,p)$ commute when they
share no link. Otherwise, they satisfy the following relations \cite{Kit97}:
\bed
A_{g_1} A_{g_2}=A_{g_1 g_2}, \hspace{0.5cm} B_{g_1} B_{g_2}=\delta_{g_1 g_2} B_{g_1},
\hspace{0.5cm} A_{g_1} B_{g_2}=B_{g_1 g_2 {g}^{-1}_1} A_{g_1},
\eed

These commutation relations represent those of an algebra called the quantum
double or Drinfeld algebra~\cite{dwp} that we denote as $D(G)$.

From the operators
\beq\label{eq:projds3}
A_s=\frac{1}{|G|} \sum_{g \in G} A_g(s), \hspace{0.3cm} \textrm{and} \hspace{0.3cm} B_p=B_e(p),
\eeq
it is possible to construct a Hamiltonian made of local mutually commuting terms, which has the same form as for the toric code, Eq.(\ref{eq:toric}). 

An elementary excitation of $H$ lives on a vertex \emph{and} an adjacent
plaquette that we call site. Each excitation is associated with a pair of
data: an equivalence class of $G$ and a representation of the normalizer of
an arbitrary element of this equivalence class \cite{dwp,Kit97,Pres:chap9}. Each
such pair of data corresponds to an irreducible representation of $D(G)$.
Quasiparticles associated with the trivial equivalence class $\{ e \}$ are
called pure charges. Quasiparticles associated with trivial representations
are called pure fluxes. The rest of the quasiparticles are called dyons. For
the $D(S_3)$ model, the quasiparticles come in eight varieties that we label
$A, \ldots, H$ \cite{Pres:chap9}. One can construct operators that project
the links related to a site onto a definite quasiparticle state. This is
done upon observing that the operators $B_h(s,p) A_g(s)$ form a reducible
representation of the quantum double $D(S_3)$. Each definite quasiparticle
state corresponds to an irreducible representation of $D(S_3)$ contained in
this reducible representation. Therefore, the projectors onto a given
quasiparticle type can be constructed using characters of the corresponding
irreducible representation
\cite{dijkgraaf,overbosch}:
\beq
P_q=\sum_{h,g \in S_3} \chi_q(h,g) B_h(s,p) A_g(s),
\eeq
where the characters $\chi_q$ associated with an irreducible representation $q$ can be calculated \cite{overbosch}.

\subsection{Pinning Quasi-Particles}

\begin{figure}[h]
\begin{center}
\includegraphics[width=40mm,height=40mm]{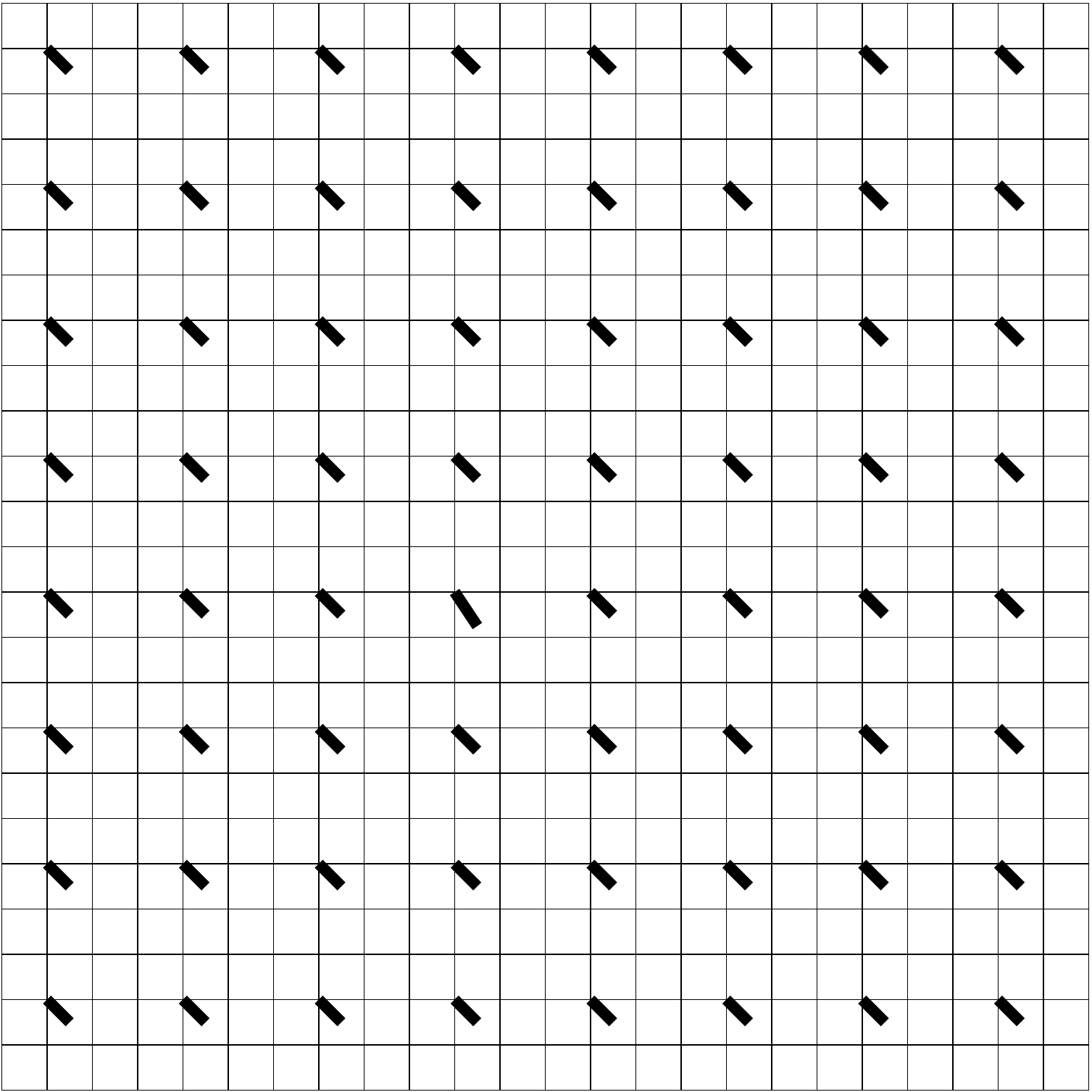}
\end{center}
\caption{Square lattice associated with a quantum double model.
Extra edges closing the lattice onto a sphere are not represented. Thick
strokes refer to sites of the truncated model where excitations, living on
a vertex and a plaquette (site), lie apart from each
other.}\label{fig:trunca}\label{fig:truncatedspectrum}
\end{figure}

We have not been able to fully characterise the spectrum of the Hamiltonian (\ref{eq:toric}), with $A_s$ and $B_p$ defined as in Eq.(\ref{eq:projds3}). However, we can argue that we actually do not need to
as far as we are interested in the topological mutual information. Excited
states of $H$ are tagged by vertices $s$ on the lattice that violate the
condition $\mean{A_s}=1$ and plaquettes $p$ that violate the condition
$\mean{B_p}=1$. We restrict to that part of the spectrum of $H$ such that
excitations are pinned at fixed, non-adjacent locations (see
Fig.~\ref{fig:trunca} for a representation of a planar region). We expect
that the more the noise undergone by the system, the less the topological
mutual information. Considering only that restricted part of the spectrum of $H$ can be
understood as additional error correction, where some plaquettes and
vertices are over-protected so that they never get excited (or only with
vanishing probability). It is actually easy to write a local Hamiltonian
that automatically does that. Therefore, we believe that the topological
mutual information of the modified model can only be larger than that of the
full spectrum.

We can also provide a renormalisation argument to support the idea of working with a truncated spectrum. Fixing the distance between two neighbour excitation sites to, say, two lattice spacings, and increasing the size of the lattice, i.e. the number of links,  we see that the truncated model has a continuum limit, where excitations are pointwise anyonic particles (or superpositions thereof). We therefore expect that the potential differences between a situation where we consider the full spectrum and a situation where we consider only the truncated spectrum vanish in the limit of large lattices.

Our third argument is that in the case of the $\mathbb{Z}_2$ toric code, it
does not make any difference whether we work with the full spectrum or with
the truncated spectrum. This is most easily seen by considering the
partition function of this model. Had we computed it for a system with more edges but still only $L^2$ possible positions for the charge (resp. flux) defects, we would have found exactly the same expression.  Similarly, one easily checks that the topological mutual information remains unchanged.

\subsection{The fusion space and partition function}

The space of $n$ excitations pinned at fixed sites has the structure \cite{Kit97}
\beq
\mathscr{H}[n]= \bigoplus_{q_1 \ldots q_n} \mathscr{H}_{q_1 \ldots q_n},
\eeq
where each index $q_i$ runs over all possible quasiparticle types $A,
\ldots, H$. This structure merely reflects the fact that different excitation patterns
lead to orthogonal states. Each space $\mathscr{H}_{q_1 \ldots q_n}$ further splits
as
\beq
\mathscr{H}_{q_1 \ldots q_n}= \mathscr{K}_{q_1} \otimes \ldots \otimes \mathscr{K}_{q_n} \otimes
\mathcal{M}_{q_1 \ldots q_n}.
\eeq
The factor spaces $\mathscr{K}_{q_i}$ have a familiar meaning. They
correspond to the local degrees of freedom of the excitations. The point of
topological models is that the extra piece $\mathcal{M}_{q_1 \ldots q_n}$,
the fusion space, may have a non-trivial dimension. The particles can
be fused. That is, two particles can be converted into one using a ribbon
operator to connect them \cite{Kit97}. In the simple Abelian case discussed
previously, the fusion rules were simple. For example, two plaquette excitations can only be
fused in a way that makes them disappear. In a general anyonic
model, the fusion rules read
\beq
q_a \times q_b= \sum_c N_{ab}^c \; q_c,
\eeq
where $N_{ab}^c$ are non-negative integers. We assume that physical
states have to fulfill some neutrality conditions, i.e. they should be
such that fusing all the particles yields with certainty the trivial
particle, denoted as $1$. This assumption is legitimate since we work
with a sphere \cite{bombin}. An important result of the theory of anyons is a formula for
the dimension of $\mathcal{M}_{q_1
\ldots q_n}$ \cite{Pres:chap9}:
\beq
\textrm{dim} \; \mathcal{M}_{q_1 \ldots q_n}=
\sum_{b_1} \ldots \sum_{b_{n-2}}
N_{q_1 q_2}^{b_1} N_{b_1 q_3}^{b_2} \ldots
N_{b_{n-2} q_n}^1.
\eeq
This formula has a structure that makes it easy to compute for arbitrary
$n$. $\textrm{dim} \; \mathcal{M}_{q_1 \ldots q_n}$ is the contraction of a (quasi) translationally 
invariant matrix product state \cite{mps}. The dimension of $\mathscr{H}[n]$ can also be calculated easily. Let $d_q$ denote the dimension associated with the quasiparticle $q$. For
$D(S_3)$ anyons, we have that
$(d_A,d_B,d_C,d_D,d_E,d_F,d_G,d_H)=(1,1,2,3,3,2,2)$ \cite{Pres:chap9}.
Defining the elements of a matrix $M$ as $M_{\alpha \beta}= \sum_{\gamma}
N_{\alpha \gamma}^{\beta} d_{\gamma}, \textrm{dim} \mathscr{H}[n] $ can be
expressed as
\beq\label{eq:totaldimension}
\textrm{dim} \mathscr{H}[n] =\sum_{q_1 \ldots q_n} \textrm{dim} \mathscr{H}_{q_1 \ldots q_n}=
\sum_{\alpha} d_{\alpha} M^{n-1}_{\alpha,1}.
\eeq
The partition function of the model on a sphere, with $n$ well separated
sites reads
\beq
Z(\beta,n)=\sum_{q_1} \ldots \sum_{q_{n}} e^{-\beta E(q_1 \ldots q_{n})}
d_{q_1} \ldots d_{q_{n}} N_{q_1 q_2}^{b_1} N_{b_1 q_3}^{b_2} \ldots N_{b_{n-2} q_{n}}^{1},
\eeq
where $E(q_1 \ldots q_{n})$ is the energy associated with a configuration
$q_1 \ldots q_{n}$. Just like $\textrm{dim} \mathscr{H}[n]$, it can be computed efficiently upon diagonalizing a matrix $M$, whose size is independent of $n$.

\subsection{von Neumann entropy at finite temperature}

Let us start by considering a scenario where a pair of anyons $\ket{q
\bar{q}}$ is created in such a way that anyon $q$ lies in some region $A$
and anyon $\bar{q}$ lies in the complementary region. In this configuration,
the von Neumann entropy of region $A$ reads \cite{Kit-Pre05}
\beq
S_q^{\textrm{pair}}(\rho_A)=S(\rho_A^{\textrm{g.s.}})+\log d_q.
\eeq

The entropy of a region when the system is in a thermal state can be
computed once we are able to calculate the entropy of a region when the
system is an arbitrary defect configuration. In turn, the latter entropy
reduces to computing the entropy when there are only \emph{two} anyons in
the system, and one lies inside the region we are interested in.

Consider a tiled sphere as the one indicated in Fig.~\ref{fig:sphere}, and
let us divide it into two simply connected regions, $A$ and $B$. Let $N_A$
and $N_B$ denote the number of sites contained in region $A$ and $B$
respectively ($N_A+N_B \equiv n$). Let $\gras{q}_A=q_1 \ldots q_{N_A}$ label
the types of anyons living on the sites contained in
region $A$ and $\gras{q}'_B=q'_1 \ldots q'_B$ label the types of anyons
living on the sites contained in region $B$. The total Hilbert space can be
decomposed as:

\beq\label{eq:decompspace}
\mathscr{H}[n]=\bigoplus_{\gras{q}_A} \bigoplus_{\gras{q}'_B}
\mathscr{K}_{\gras{q}_A} \otimes \mathscr{K}_{\gras{q}'_B}
\otimes
\bigoplus_b \mathcal{M}^b_{\gras{q}_A} \otimes \mathcal{M}^{\bar{b}}_{\gras{q}'_B}
\otimes \mathcal{M}^1_{b,\bar{b}}.
\eeq
(N.B. The spaces $\mathcal{M}^1_{b,\bar{b}}$ are one-dimensional.) The decomposition (\ref{eq:decompspace}) induces the following representation of the thermal
state of the $D(S_3)$ superconductor
\bed
\rho_{\textrm{th}}=
\bigoplus_{\gras{q}_A} \bigoplus_{\gras{q}'_B}
\bigoplus_{b=A}^{H}
\bigoplus_{\mu_1=1}^{ \textrm{dim} \mathcal{M}^b_{\gras{q}_A} }
\bigoplus_{\mu_2=1}^{ \textrm{dim} \mathcal{M}^{\bar{b}}_{\gras{q}'_B}}
\bigoplus_{s_A=1}^{d(\gras{q}_A)}
\bigoplus_{s_B=1}^{d(\gras{q'}_B)}
\frac{e^{-\beta (E(\gras{q}_A)+E(\gras{q}'_B))}}{Z(\beta,n)}
\eed
\beq\label{eq:decompthermalnonAbelian}
\ket{\gras{q}_A \to b,\mu_1, s_A; \gras{q}'_B \to \bar{b},\mu_2,s_B}
\bra{\gras{q}_A \to b,\mu_1, s_A; \gras{q}'_B \to \bar{b},\mu_2,s_B},
\eeq
where $E(\gras{q}_A)=\sum_{j=1}^{N_A} E(q_j)$,
$d(\gras{q}_A)=\prod_{j=1}^{N_A} d(q_j)$, $\mu_1$ (resp. $\mu_2$) denotes
the possible channels through which the anyons $\gras{q}_A$ (resp.
$\gras{q'}_B$) can be fused into $b$ (resp. $\bar{b}$), and $s_A$ (resp.
$s_B$) is a collective index for the internal degrees of freedom of the
quasiparticles contained in region $A$ (resp. $B$).

Defining
\bed
Z_A(\beta,b) \equiv\sum_{\gras{q}_A} d(\gras{q}_A) e^{-\beta E(\gras{q}_A)}
\textrm{dim} \mathcal{M}^{b}_{\gras{q}_A},
\eed
and $Z_B(\beta,\bar{b})$ likewise, we get that the reduced state of region $A$ reads

\beq\label{eq:reducedthermalnonAbelian}
\rho_A=\bigoplus_{b,\gras{q}_A}
\bigoplus_{\mu_1=1}^{ \textrm{dim} \mathcal{M}^b_{\gras{q}_A}}
\bigoplus_{s_A=1}^{d(\gras{q}_A)}
\frac{e^{-\beta E(\gras{q}_A)} Z_B(\beta,\bar{b}) }{Z(\beta,n)}
\tr_B
\ket{\gras{q}_A \to b,\mu_1, s_A; \bar{b}}
\bra{\gras{q}_A \to b,\mu_1, s_A; \bar{b}}.
\eeq
It is worth observing the resemblance between
Eq.(\ref{eq:reducedthermalnonAbelian}) and the reduced density matrix for
the toric code. From Eq.(\ref{eq:reducedthermalnonAbelian}) and
Eq.(\ref{eq:entropydirectsum}), we deduce an expression for the entropy of
system $A$:
\beq
S(\rho_A)=
-\sum_{\gras{q}_A,b,\mu_1,s_A} \frac{e^{-\beta E(\gras{q}_A)} Z_B(\beta,\bar{b}) }{Z(\beta,n)}
\Big\{
\ln \big[\frac{e^{-\beta E(\gras{q}_A)} Z_B(\beta,\bar{b}) }{Z(\beta,n)}
\big]
-S_q^{\textrm{pair}}(\rho_A)
\Big\}
\eeq
Just as for the partition function, $S(\rho_A)$ can be computed by
diagonalization of an auxiliary matrix. Finally, the
von Neumann entropy of the whole sphere can be efficiently computed from the
partition function using again Eq.(\ref{eq:Zentropy}).

These expressions are valid for any quantum double model on a lattice such as the one depicted in Fig.~\ref{fig:truncatedspectrum}. Remarkably, the only microscopic data involved in the expressions for $S(\rho_A)$ and $S(\rho_{\textrm{th}})$ are the tensor $N$ governing the fusion rules, the energies associated with each quasiparticle type and their internal dimension. It is therefore tempting to conjecture that these expressions are valid for other anyonic theories. 

The fact that the fusion rules tensor $N_{**}^*$ appears in the expression for the finite temperature entropy of a region marks a qualitative difference with respect to the zero temperature (ground state) case. This is interesting in that it shows that entropic quantities could be used in order to discriminate between different topological models. For example, at finite temperature, the quantum double models based on $\mathbb{Z}_6$ and $S_3$ exhibit different region entropies, while these quantities coincide at zero temperature. We understand this difference as follows: at finite temperature, quasiparticle pairs can be spontaneously created from the vacuum and can move around on the lattice. When they braid, they give rise to qualitatively different statistics, which in turn give rise to different values for entropic quantities.

\subsection{Numerical Results}

The results derived in the previous section, and a rigorous calculation of
the ground state entropy, shown in Appendix \ref{sec:sgs} have allowed us to
study numerically how the topological mutual information behaves as a
function of $\beta$. The systems we have considered are four tiled spheres,
all with $96 \times 96$ plaquettes. The first sphere contains $64$ sites,
the second $144$, the third $256$ and the fourth $576$. Although these
systems are small when compared to their Abelian counterpart we have
considered on Fig.\ref{fig:puretopentropy}, they are large enough to show that non-Abelian systems
are affected by temperature in the same way as the toric code \cite{Iblisdir}. 

\begin{figure}[h]
\begin{center}
\includegraphics[width=60mm,height=45mm]{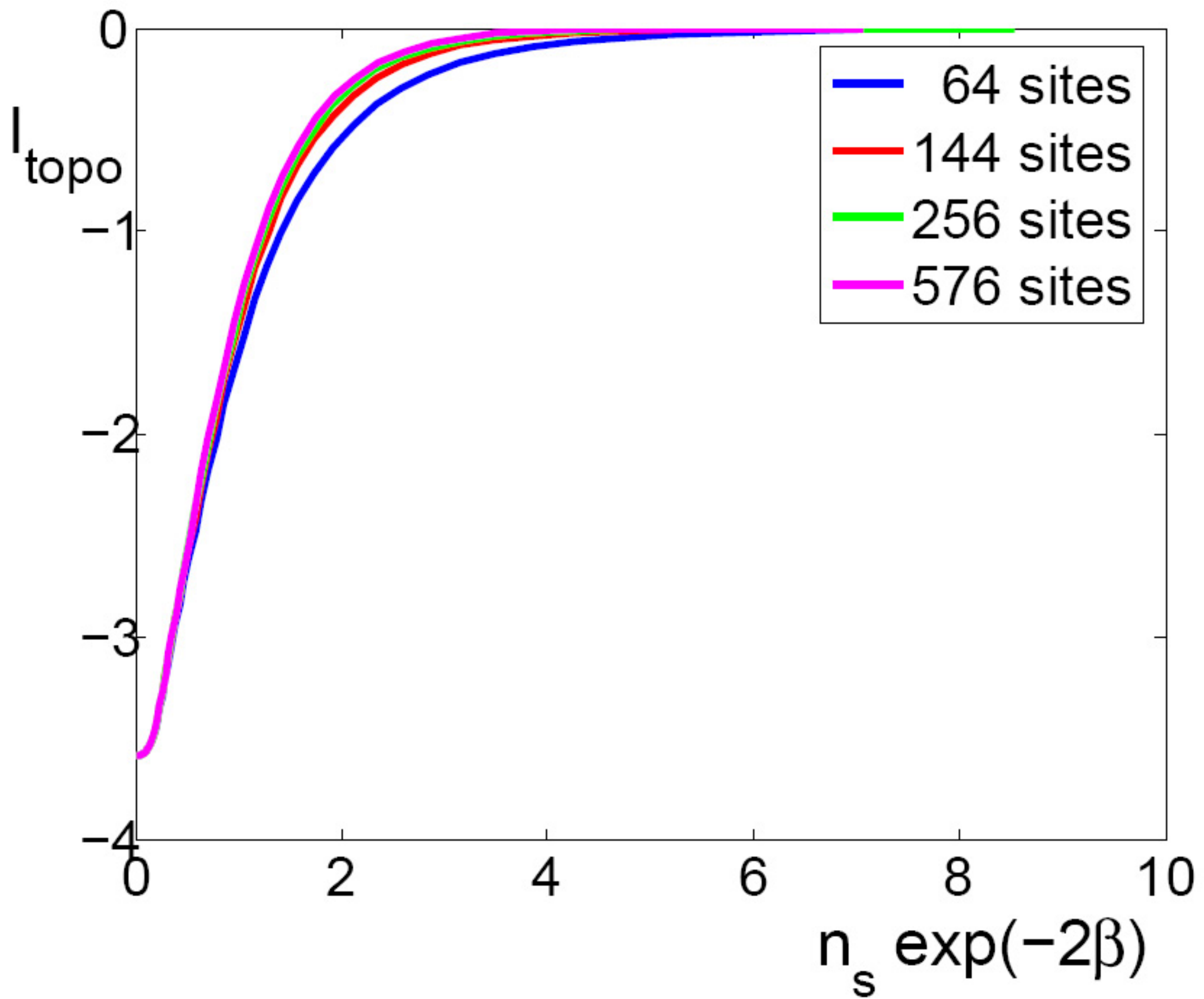}
\caption{Topological mutual information as a function of $n \exp(-2\beta)$ for the $D(S_3)$ model.}\label{fig:narescaleddata}
\end{center}
\end{figure}

Fig.~\ref{fig:narescaleddata} is presented to analyze the existence of scaling laws in the non-Abelian
case. There, the topological mutual information is represented as a function of $n
\exp(-2\beta)$. We can see that the curves tend to collapse when the number of sites is
increased, that is, the distance between the four curves reduces. This fact
supports the idea that for large systems, the topological mutual information
only depends on the `volume' of the system and the temperature through the
product $n \exp(-2\beta)$. This result is consistent with the scaling relation found for the toric code.

\section{Conclusions}\label{sec:conclusions}

In conclusion, we have studied how topological phases of lattice systems behave as a function of temperature, using the topological mutual information as an order parameter.  We have focused on two
specific models; the toric code and the $D(S_3)$ quantum double model. These systems turned out to be affected by temperature in much the same way. We have studied how the topological mutual information depends on size and temperature through a simple scaling variable. In particular, we have shown how an increase of system size can be compensated by a vanishing decrease of temperature. It would be interesting to see whether our findings are also valid for other systems, such as a FQHE sample for instance. We have also seen how temperature could allow to resolve between models whose topological entropy are equal at zero temperature. We have also introduced a class of simulations where running time simulates increasing temperature.  Though exemplified by a minimal plaquette in the Abelian model, it can be extended to more complicated situations. We have also discussed why, at finite temperature, $I_{\textrm{topo}}$ should be used as an appropriate topological order parameter instead of the topological entropy. We have shown how $I_{\mathrm{topo}}$ relates directly to the thermal probability distribution of topological charges, and how in turn this distribution is related to the mean value of $S$-matrix elements and controls the visibility of anyonic processes, and therefore the usefulness of topologically ordered systems as quantum memories. 

\section{Acknowledgements}

We thank Gavin Brennen, Ignacio Cirac, Jos\'e Ignacio Latorre, Miguel-Angel Martin-Delgado, Ady
Stern and Paolo Zanardi for inspiring discussions. We also thank Miguel-Angel 
Martin-Delgado for having given us an early version of Ref.~\cite{bombin}.
Financial support from the Generalitat de Catalunya, MEC (Spain), QAP,
EMALI, SCALA (EU), Spanish grants MTM2005-00082, I-MATH, and
CCG07-UCM/ESP-2797, and UK grants from EPSRC and the Royal Society.

\appendix

\section{Some calculations for the toric code}\label{sec:detailstoric}

Most of the calculations presented in this appendix are based on repeated use of three elementary identities:
\bed
\sum_{k=0}^N \binom{N}{k} \alpha^k=(1+\alpha)^N,
\eed
\bed
\sum_{ \substack{ k=0 \\  k  \; \textrm{even} }}^N \binom{N}{k} \alpha^k= \frac{1}{2} \sum_{k=0}^N \binom{N}{k} (\alpha^k+(-\alpha)^k),
\eed
\bed
\sum_{k=0}^N \binom{N}{k} k \; \alpha^k=\alpha \frac{\partial}{\partial \alpha} (1+\alpha)^N=N \; \alpha (1+\alpha)^{N-1}.
\eed

One finds that the partition function reads:
\beqa
Z(\beta,L) &=& \sum_{i=0}^{L^2}e^{-\beta E_i} d_i=
\sum_{n_\phi=0}^{\lfloor L^2/2 \rfloor}
\sum_{n_c=0}^{\lfloor L^2/2 \rfloor}
e^{-\beta(E_0+ 4 n_\phi+4 n_c)}
\binom{L^2}{2 n_\phi}
\binom{L^2}{2 n_c} \nonumber \\
&=& [(2 \cosh \beta)^{L^2}+(2 \sinh \beta)^{L^2}]^2,
\eeqa
as derived in \cite{Cast06,nussinov}.

Next we want to compute the quantity
\beq
W(\beta,R,L) \equiv -\sum_{\gras{q}_R} 4 C(\gras{q}_R) \ln(4 C(\gras{q}_R)),
\eeq
appearing in (\ref{eq:entropyR}). For that, it is convenient to consider plaquette and vertex excitations separately: $\gras{q}_R \equiv (\phi_R, c_R)$. Let $N_p(R)$ (resp. $N_*(R)$) denote the number of plaquette (resp. vertex) excitations whose edges are all contained in $R$, and let us define $N'_p(R) \equiv L^2- N_p(R)$ and $N'_*(R)=L^2-N_*(R)$. With these notations, $C(\gras{q}_R)$ can be decomposed as
\bed
C(\gras{q}_R)=\frac{e^{-\beta(E_0+2 |\phi_R|+ 2 |c_R|)}}{Z(\beta,L)} \;
\mathscr{P}(|\phi_R|,N' _p(R),e^{-2 \beta})\;
\mathscr{P}(|c_R|,N' _*(R),e^{-2 \beta}),
\eed
where
\beq
\mathscr{P}(m,N,\alpha)=\sum_{\substack{ n=0 \\ n+m \; \textrm{even}}}^N \binom{N}{n} \alpha^n=
\frac{1}{2} \{ (1+\alpha)^N+ (-1)^m (1-\alpha)^N  \}.
\eeq
In order to get an anlytical expression for $W(\beta,R,L)$, it is sufficient to evaluate sums like:
\bed
S_1(R) =\sum_{\phi_R} e^{-2 \beta |\phi_R|} \mathscr{P}(|\phi_R|, N'_p(R),e^{-2\beta}),
\eed
and
\bed
S_2(R) = \sum _{\phi_R} e^{-2 \beta |\phi_R|} \mathscr{P}(|\phi_R|, N'_p(R),e^{-2\beta})
\ln \mathscr{P}(|\phi_R|, N'_p(R),e^{-2\beta}).
\eed
\beqa
S_1(R) &=&  \sum_{\phi_R} e^{-2 \beta |\phi_R|} \mathscr{P}(|\phi_R|,N'_p(R),e^{-2\beta})
\nonumber \\
&=& \frac{1}{2} \sum_{m=0}^{N_p(R)} e^{-2 \beta m} \binom{N_p(R)}{m} \{ (1+e^{-2\beta})^{N'_p(R)}
+(-)^m (1-e^{-2\beta})^{N'_p(R)} \} \nonumber\\
&=& \frac{1}{2} \{  (1+e^{-2\beta})^{L^2}+(1-e^{-2\beta})^{L^2}\}.
\eeqa
Splitting $S_2(R)$ as a sum over even values of $|\phi_R|$ and a sum over odd values of $|\phi_R|$, we get that
\bed
S_2(R)=\frac{1}{4} \{ (1+e^{-2 \beta})^{N_p(R)} + (1-e^{-2 \beta})^{N_p(R)}  \}
 \{ (1+e^{-2 \beta})^{N'_p(R)} + (1-e^{-2 \beta})^{N'_p(R)}  \}
\eed
\bed
 \ln [\frac{1}{2} \{ (1+e^{-2 \beta})^{N'_p(R)} + (1-e^{-2 \beta})^{N'_p(R)}   \}] \\
\eed
\bed
+\frac{1}{4} \{ (1+e^{-2 \beta})^{N_p(R)} - (1-e^{-2 \beta})^{N_p(R)}  \}
 \{ (1+e^{-2 \beta})^{N'_p(R)} - (1-e^{-2 \beta})^{N'_p(R)}  \}
\eed
\beq
\ln [\frac{1}{2} \{ (1+e^{-2 \beta})^{N'_p(R)} - (1-e^{-2 \beta})^{N'_p(R)}   \}]
\eeq
With these results, we find Eqs.(\ref{eq:entropyfinitet}-\ref{eq:entropytorus}).

\section{Rough edges and smooth edges}

\begin{figure}[h]
\begin{center}
\includegraphics[width=90mm,height=28mm]{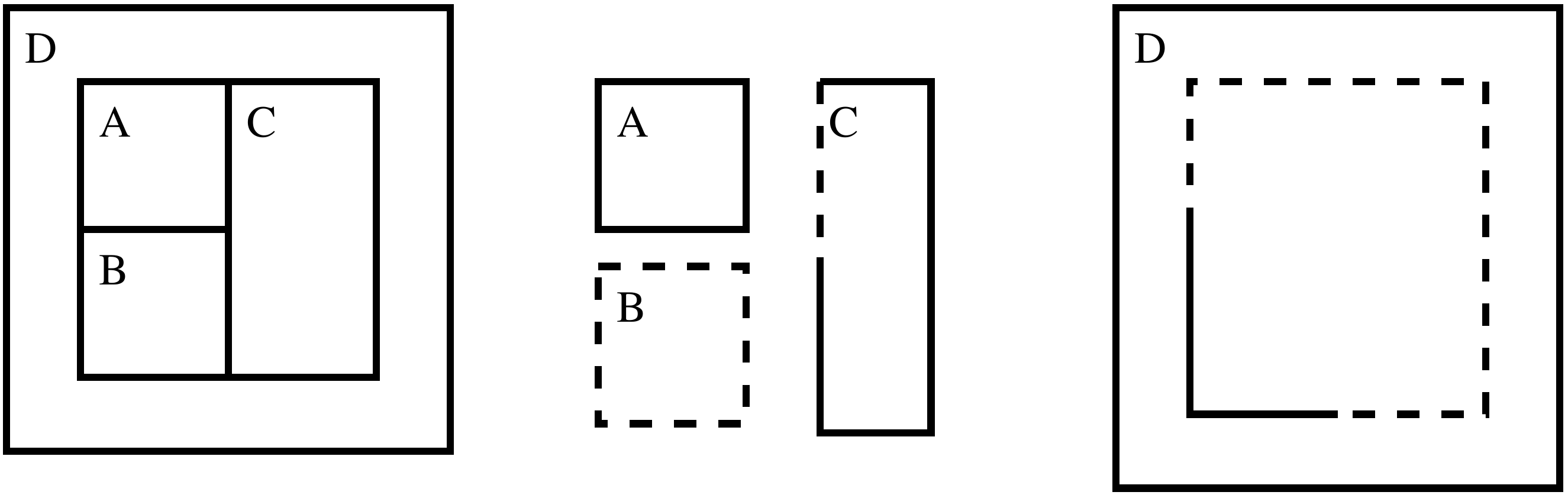}
\caption{Division of the lattice into four regions used in numerical calculations. Delimitations of a region by dotted lines indicate that the corresponding links do not belong to the region.}
\label{fig:roughsmooth}
\end{center}
\end{figure}

When dividing a lattice into four regions as indicated in
Fig.~\ref{fig:torusdivided}, one has to be careful that the links on the
boundary between two regions should be attributed only to one of them. On Fig.\ref{fig:roughsmooth}, we show how we have chosen to divide Fig.\ref{fig:torusdivided}. Data related to the division are collected in Table 1.

In the Abelian case, the entropy of a region $R$ in a zero temperature thermal state (i.e. in an equal mixture of four basis states of the ground states) is given by $(|\partial_R|-1)\ln 2-\epsilon_R$. The $\epsilon_R$ correction appears because we have worked with a model defined on a torus, and takes into account the fact that the region $R$ might be contractible. The ``area" $|\partial R|$ is given by the total number of crosses minus the number of crosses fully inside $R$ minus the number of crosses fully outside
$R$.

The non-Abelian $D(S_3)$ case has been studied on a sphere. There the ground state entropy is given by $(|\partial_R|-1) \ln 6$, see Appendix C. As shown in Fig.~\ref{fig:sphere}, there is a region within the sphere made of
square plaquettes only. (We have chosen our tiling of the sphere in this way.) Let $L$ denote the (linear) size of this region, and $k$ denotes the linear size of region $A$. (All triangular plaquettes are assumed to be inside region $D$.) In order to ease the computations, we choose $L$ to be a multiple of $k$, and we impose that the density of sites, $s^2$, is such that $s^{-1}$ is a divisor of $k$. (Triangular plaquettes are assumed to support no site.)

\begin{table}[h]
\begin{tabular}{|c|c|c|c|c|c|}
\hline
Region  & ``Area"-1 & $\epsilon$ & $N_p$ & $N_*$ & ``Volume" \\
\hline
A & $4k-1$ & $-\log 4$ & $k^2$ & $(k-1)^2$ & $k^2$ \\
B & $4k-5$ & $-\log 4$ & $k^2-4k+4$ & $(k-1)^2$ & $k^2$ \\
C & $6k-1$ & $-\log 4$ & $2 k^2-k$ & $(k-1)(2k-1)$ & $2k^2$ \\
D & $8k-2$ & $0$ & $L^2-4k^2-6k$ & $L^2-4 k^2-4k$ & $L^2-4k^2$ \\
AB & $6k-3$ & $-\log 4$ & $2k^2-3k+2$ & $(k-1)(2k-1)$ & $2k^2$ \\
AC & $8k-2$ & $-\log 4$ & $3k^2$ & $3 k^2-4k+2$ & $3k^2$ \\
AD & $8k-3$ & $0$ & $L^2-3k^2-4k$ & $L^2-3 k^2-4k+1$ & $L^2-3k^2$ \\
BC & $8k-3$ & $-\log 4$ & $3k^2-4k+2$ & $3 k^2-4k+1$ & $3 k^2$ \\
BD & $8k-2$ & $0$ & $L^2-3k^2-8k+1$ & $L^2-3 k^2-4k-1$ & $L^2-3k^2$ \\
CD & $6k-3$ & $0$ & $L^2-2k^2-3k$ & $L^2-2 k^2-3k+1$ & $L^2-2k^2$ \\
ABC & $8k-2$ & $-\log 4$ & $4k^2-2k+1$ & $(2k-1)^2$ & $4 k^2$ \\
ABD & $6k-1$ & $0$ & $L^2-2k^2-5k$ & $L^2-2 k^2-3k-1$ & $L^2-2k^2$ \\
ACD & $4k-5$ & $0$ & $L^2-k^2$ & $L^2-k^2-2k+3$ & $L^2-k^2$ \\
BCD & $4k-1$ & $0$ & $L^2-k^2-4k$ & $L^2-k^2-2k-1$ & $L^2-k^2$ \\

\hline

\end{tabular}
\caption{Some data useful to compute the von Neumann entropies of various regions of a tiled torus at finite temperature. The regions are those depicted in Fig.~\ref{fig:roughsmooth}.
$\epsilon$ labels the correction to the ground state entropy due to the
topology of the considered region. $N_p$ (resp. $N_*$) denotes the number of
plaquettes (resp. crosses) fully contained in the region.}
\end{table}\label{tab:dataforentropy}

\section{Ground State Entanglement of the $D(S_3)$ superconductor}\label{sec:sgs}

Here, we generalize some of the calculations presented in
\cite{hamma:bilocal} and derive a simple exact expression for the entropy of
the reduced matrix $\rho_A$ of the ground state $|\xi\>$ in a region $A$.
($B$ here denotes the complementary of $A$). We need to distinguish amongst three
types of vertices, those in $A$, those in $B$ and those touched by edges
belonging to $A$ and edges belonging to $B$. The set of vertices of the
latter kind will be referred to as $\partial A$. We consider only regions
with the following properties
\begin{enumerate}

\item Both $A$ and $B$ are connected in the sense of vertices. That is, every two vertices inside $A$ (resp. $B$) can be connected by a path of adjacent vertices inside $A$ (resp. $B$).

\item For each $s\in\partial A$, there exist $s'\in A, s''\in B$ adjacent to $s$.

\end{enumerate}

The (unnormalized) ground state of the $D(S_3)$ model defined on a sphere can be written as
$$|\xi\>=\sum_{g_1,\ldots, g_N} A_{g_1}(s_1)\cdots A_{g_N}(s_N)|e\cdots e\>=\sum_{\gras{g} \in \gras{G}} \pi(\gras{g}) |e\cdots e\>,$$
where $\gras{g} \equiv (g_1,\ldots, g_N)$, $\gras{G}=S_3\times \cdots \times S_3$ and $\pi$ is the
representation of $\gras{G}$ defined by $\pi(\gras{g})=A_{g_1}(s_1)\cdots
A_{g_N}(s_N)$. The action of $\pi(\gras{g})$ can be decomposed as a product of unitaries acting on the edges of the lattice:

\bed
\pi(\gras{g})= \bigotimes_{\langle s_i, s_j \rangle} U_{\langle s_i, s_j \rangle}(g_i,g_j),
\eed
where the operator $U_{\langle s_i, s_j \rangle}$ acts as follows on the
`computational' basis states $\{ \ket{x}: x \in S_3 \}$ of the edge $\langle
s_i, s_j \rangle$: $U_{\langle s_i, s_j \rangle} \ket{x}=
\ket{g_i x g_j^{-1}}$ if the edge goes from the vertex $s_i$ to $s_j$. Therefore $U_{\langle s_i, s_j \rangle}(g_i,g_j)$ acts trivially on an edge $\langle s_i, s_j \rangle$ only if  $g_i$ equals $g_j$ and belongs to the centraliser of $S_3$, which is $\{e\}$. Therefore, the representation $\pi$ is faithful and we can (and will) identify $\gras{G}$ with $\pi(\gras{G})$.

We now borrow some definitions from \cite{hamma:bilocal}. We define two normal subgroups in $\gras{G}$:
\bed
\gras{G}_A=\{ \gras{g} \in \gras{G} | g_j=e \text{ if } s_j\not \in A \},
\eed
\bed
\gras{G}_B=\{ \gras{g} \in \gras{G} | g_j=e \text{ if } s_j\not \in B \},
\eed
and we call $\gras{G}_{AB}$ the quotient group $\gras{G} / (\gras{G}_A \times \gras{G}_B)$.

Any element of $\gras{G}_X$ ($X=A,B$) acts only on $X$ and we have the following partition of $\gras{G}$:

$$\gras{G}=\cup_{[\gras{h}]\in \gras{G}_{AB}}\{(\gras{g}_A\otimes \gras{g}_B) \gras{h} | \gras{g}_A\in \gras{G}_A, \gras{g}_b\in \gras{G}_B\},$$
where we can fix any representative $\gras{h}$ for the class $[\gras{h}]$.  This decomposition implies that the ground state can be expressed as
\beq\label{eq:xidecompcoset}
|\xi\>= Q_A\otimes Q_B \sum_{[\gras{h}]} \gras{h} |e\cdots e\>,
\eeq
where $Q_X=\sum_{\gras{g}_X \in \gras{G}_X} \gras{g}_X$.

For our purposes, we need to quotient further $G_{AB}$. So, we consider the
diagonal of $\gras{G}$, $\gras{G}_{\textrm{d}}=\{\tilde{r}=(r,\cdots,r)
|r\in S_3\}$, and  the quotient set, $\gras{G}_{AB}/\gras{G}_{\textrm{d}}$,
defined by the (right) equivalence relation: $[\gras{h}] \sim [\gras{h}']$
iff $\exists \; r \in S_3$ s.t. $[\gras{h}] \sim [\gras{h}' \tilde{r}]$.
Clearly, $|\gras{G}_{AB}/\gras{G}_{\textrm{d}}|=|\gras{G}_{AB}|/|S_3|$. We
denote $[[\gras{h}]]$ the elements of $\gras{G}_{AB}/\gras{G}_{\textrm{d}}$.
Since $\gras{G}_{\textrm{d}}$ acts trivially on $\ket{e \ldots e}$,
$\ket{\xi}$ can be further decomposed as

\beq\label{eq:xidecompsupercoset}
\ket{\xi}= Q_A \otimes Q_B \sum_{r \in S_3} \sum_{[[\gras{h}]]} \gras{h} \; \tilde{r} \ket{e \ldots e}=
|S_3| Q_A \otimes Q_B  \sum_{[[\gras{h} ]]} \gras{h}_A \otimes \gras{h}_B \ket{e \ldots e},
\eeq
with $\gras{h}=\gras{h}_A \otimes \gras{h}_B$, and where $\gras{h}_A$ (resp.
$\gras{h}_B$) does not necessarily belong to $\gras{G}_A$ (resp.
$\gras{G}_B$). In order to conclude our calculation of $S(\rho_A)$, we use
the following lemma.

\begin{lemma}\label{lem:gsentropy}

Let $\gras{g}$ denote an arbitrary element of $\gras{G}$ and let $\gras{g}=\gras{g}_A \otimes \gras{g}_B$ denote its decomposition into an operator acting on $A$ and an operator acting on $B$. If for all $r \in S_3$, $[\gras{g}] \neq [\tilde{r}]$, then $\bra{e \ldots e} \gras{g}_X \ket{e \ldots e}=0$ for both $X=A,B$.
\end{lemma}

\emph{Proof}: Let us suppose that it is false. $[\gras{g}] \neq [\tilde{r}]$ iff there are at least two vertices $s_i,s_j \in \partial A$ such that $g_i \neq g_j$. Let us consider a path $\Gamma_X$ connecting $s_i$ to $s_j$ through edges within $X$. Since
\bed
\bra{e \ldots e} \gras{g}_X \ket{e \ldots e}= \prod_{\langle s_{\alpha}, s_{\beta \rangle} \in X}
\bra{e} g_{\alpha}^{-1} g_{\beta} \ket{e},
\eed
we have that $\bra{e \ldots e} \gras{g}_X \ket{e \ldots e} \neq 0$ only if $g_{\alpha}=g_{\beta}$ for all pairs of adjacent vertices $s_{\alpha},s_{\beta}$. Thus $\bra{e \ldots e} \gras{g}_X \ket{e \ldots e} \neq 0$ only if $g_i=g_{\alpha}$ for all vertex $s_{\alpha} \in \Gamma_X$, and in particular $g_i=g_j$, which is the desired contradiction.

This lemma allows us to prove that the states $\{ (\gras{h}_A \otimes \gras{h}_B) \ket{e \ldots e}: [[\gras{h}]] \in \gras{G}_{AB}/\gras{G}_{\textrm{d}} \}$ form a bi-orthogonal set, so that the expression
(\ref{eq:xidecompsupercoset}) is actually a Schmidt decomposition of the ground state. If
$[[ \gras{h} ]]$ and $[[ \gras{h}' ]]$ are different equivalence classes, then $\bra{e \ldots e} \gras{h}^{-1}_X \gras{h}^{'}_X \ket{e \ldots e}=0$ for both $X=A,B$. Indeed, $[[ \gras{h} ]] \neq [[ \gras{h}' ]]$ means that $[\gras{h} ] \neq [ \gras{h}' \tilde{r} ]$ for all $r \in S_3$. That is $[\gras{h}^{-1} \gras{h}' ] \neq [\tilde{r}]$ and the lemma \ref{lem:gsentropy} shows that the desired property holds. At this point, the construction of Ref.\cite{hamma:bilocal} can be used to first prove that the set $\{ (Q_A \gras{h}_A \otimes Q_B \gras{h}_B) \ket{e \ldots e}: [[\gras{h}]] \in \gras{G}_{AB}/\gras{G}_d \}$ is also a bi-orthogonal set, and then on to show that
\beq
S(\rho_A)=\log_2 \frac{|\gras{G}_{AB}|}{|S_3|}=\log_2 |S_3| (N_{\partial A}-1),
\eeq
where $N_{\partial A}$ is the number of vertices on the boundary of $A$.

\emph{Remark:} The argument just presented for $S_3$ is actually valid for any finite group $G$. If $G$ has a non-trivial centre $Z(G) \neq \{ e\}$, then one would merely start the construction from the group $\gras{G}/Z(\gras{G})_{\textrm{d}}$ instead of $\gras{G}$.

\end{document}